\documentclass[preprint]{ptephy}

\preprintnumber{KYUSHU-HET-206}

\usepackage{amssymb}
\usepackage{amsthm}
\usepackage{amsmath}
\usepackage{booktabs}
\usepackage{bbm}
\usepackage{mathtools}
\usepackage{subcaption}

\DeclareMathOperator{\Det}{Det}

\newcommand{\Slash}[1]{{\ooalign{\hfil/\hfil\crcr$#1$}}}
\numberwithin{equation}{section}

\usepackage{hyperref}

\allowdisplaybreaks[3]

\begin{document}

\title{Vacuum energy of the supersymmetric $\mathbb{C}P^{N-1}$ model on
$\mathbb{R}\times S^1$ in the $1/N$ expansion}

\author{%
\name{\fname{Kosuke} \surname{Ishikawa}}{1},
\name{\fname{Okuto} \surname{Morikawa}}{1},
\name{\fname{Kazuya} \surname{Shibata}}{1},
and
\name{\fname{Hiroshi} \surname{Suzuki}}{1,\ast}
}

\address{%
\affil{1}{Department of Physics, Kyushu University
744 Motooka, Nishi-ku, Fukuoka, 819-0395, Japan}
\email{hsuzuki@phys.kyushu-u.ac.jp}
}

\date{\today}

\begin{abstract}
By employing the $1/N$ expansion, we compute the vacuum
energy~$E(\delta\epsilon)$ of the two-dimensional supersymmetric (SUSY)
$\mathbb{C}P^{N-1}$ model on~$\mathbb{R}\times S^1$ with $\mathbb{Z}_N$ twisted
boundary conditions to the second order in a SUSY-breaking
parameter~$\delta\epsilon$. This quantity was vigorously studied recently by
Fujimori et\ al.\ using a semi-classical approximation based on the bion,
motivated by a possible semi-classical picture on the infrared renormalon. In
our calculation, we find that the parameter~$\delta\epsilon$ receives
renormalization and, after this renormalization, the vacuum energy becomes
ultraviolet finite. To the next-to-leading order of the $1/N$ expansion, we
find that the vacuum energy normalized by the radius of the~$S^1$, $R$,
$RE(\delta\epsilon)$ behaves as inverse powers of~$\Lambda R$ for~$\Lambda R$
small, where $\Lambda$ is the dynamical scale. Since $\Lambda$ is related to
the renormalized 't~Hooft coupling~$\lambda_R$
as~$\Lambda\sim e^{-2\pi/\lambda_R}$, to the order of the $1/N$ expansion we work
out, the vacuum energy is a purely non-perturbative quantity and has no
well-defined weak coupling expansion in~$\lambda_R$.
\end{abstract}

\subjectindex{B06, B16, B32, B34, B35}
\maketitle

\section{Introduction}
\label{sec:1}
In this paper, by employing the $1/N$ expansion (for a classical exposition,
see~Ref.~\cite{Coleman:1985rnk}), we compute the vacuum
energy~$E(\delta\epsilon)$ of the two-dimensional (2D) supersymmetric (SUSY)
$\mathbb{C}P^{N-1}$ model~\cite{Cremmer:1978bh,DAdda:1978dle,Witten:1978bc}
on~$\mathbb{R}\times S^1$ with $\mathbb{Z}_N$ twisted boundary conditions to
the second order in a SUSY-breaking parameter~$\delta\epsilon$. This quantity
was vigorously studied recently by Fujimori et\ al.~\cite{Fujimori:2018kqp}
(see also~Refs.~\cite{Fujimori:2016ljw,Fujimori:2017oab,Fujimori:2017osz})
using a semi-classical approximation based on the bion~\cite{Eto:2004rz,%
Eto:2006mz,Eto:2006pg,Bruckmann:2007zh,Brendel:2009mp,Bruckmann:2018rra}. One
of the motivations for their study was a possible semi-classical picture on the
infrared (IR) renormalon~\cite{tHooft:1977xjm,Beneke:1998ui} advocated
in~Refs.~\cite{Argyres:2012vv,Argyres:2012ka,Dunne:2012ae,Dunne:2012zk}. In
these works, in the context of the resurgence program (for a review,
see~Ref.~\cite{Dunne:2016nmc} and the references cited therein), it is proposed
that the ambiguity caused by the IR renormalon through the Borel resummation
(for a review, see~Ref.~\cite{LeGuillou:1990nq}) be cancelled by the ambiguity
associated with the integration of quasi-collective coordinates of the bion;
this scenario is quite analogous to the Bogomolny--Zinn-Justin mechanism for
the instanton--anti-instanton pair~\cite{Bogomolny:1980ur,ZinnJustin:1981dx}.

In~Ref.~\cite{Fujimori:2018kqp}, by using the Lefschetz thimble
method~\cite{Witten:2010cx,Cristoforetti:2012su,Fujii:2013sra}, the integration
over quasi-collective coordinates of the bion is explicitly carried out and it
was found that the vacuum energy~$E(\delta\epsilon)$ possesses the imaginary
ambiguity which is of the same order as that caused by the so-called $u=1$ IR
renormalon. On the other hand, for the four-dimensional $SU(N)$ gauge theory
with the adjoint fermion (4D QCD(adj.)), for $N=2$ and~$3$, it has been
found~\cite{Anber:2014sda} that when the spacetime is compactified
as~$\mathbb{R}^3\times S^1$, the logarithmic behavior of the vacuum
polarization of the gauge boson associated with the Cartan subalgebra
(``photon'') disappears. Since the IR renormalon is attributed to such a
logarithmic behavior, in~Ref.~\cite{Anber:2014sda} it is concluded that the
circle compactification generally eliminates the IR renormalon. This appears
inconsistent with the renormalon interpretation of the result
in~Ref.~\cite{Fujimori:2018kqp}.

The original motivation in a series of
works~\cite{Ishikawa:2019tnw,Ashie:2019cmy,Ishikawa:2019oga} by a group
including the present authors was to investigate the fate of the IR renormalon
under the circle compactification to understand the above
inconsistency.\footnote{Recent related works are
Refs.~\cite{Yamazaki:2019arj,Marino:2019fvu,Flachi:2019yci}.} For this, we
employed the $1/N$ expansion (i.e.\ the large-$N$ limit), in which
\begin{equation}
   \text{$\Lambda R=\text{const.}$ as $N\to\infty$},
\label{eq:(1.1)}
\end{equation}
where $\Lambda$ is a dynamical scale and $R$ is the $S^1$ radius. We expected
that in this way the IR renormalon and the bion can be highlighted, because the
beta function of the 't~Hooft coupling and the bion action remain non-trivial
in the large-$N$ limit, Eq.~\eqref{eq:(1.1)}, whereas other sources to the
Borel singularity such as the instanton--anti-instanton pair are suppressed.
This intention was not so successful, because the calculations
in~Refs.~\cite{Ishikawa:2019tnw,Ashie:2019cmy,Ishikawa:2019oga} show that the
behavior of the IR renormalon rather depends on the system; in the 2D SUSY
$\mathbb{C}P^{N-1}$ model, the compactification from~$\mathbb{R}^2$
to~$\mathbb{R}\times S^1$ shifts the location of the Borel singularity
associated with the IR renormalon~\cite{Ishikawa:2019tnw,Ishikawa:2019oga}. In
the 4D QCD(adj.), because of the twisted momentum of the gauge boson associated
with the root vectors (``W boson''), $\mathbb{R}^3\times S^1$ is effectively
decompactified in the large-$N$
limit~\cite{Eguchi:1982nm,Gross:1982at,Sulejmanpasic:2016llc} and the IR
renormalon gives rise to the same Borel singularity as the
uncompactified~$\mathbb{R}^4$~\cite{Ashie:2019cmy}.\footnote{In this analysis,
we relied on the so-called large-$\beta_0$
approximation~\cite{Beneke:1994qe,Broadhurst:1993ru,Ball:1995ni}.} It appears
that a unified picture on the semi-classical understanding of the IR renormalon
is still missing.

In the present paper, as announced in~Ref.~\cite{Ishikawa:2019tnw}, in the
$1/N$ expansion with~Eq.~\eqref{eq:(1.1)}, we compute the vacuum
energy~$E(\delta E)$ of the 2D SUSY $\mathbb{C}P^{N-1}$ model
on~$\mathbb{R}\times S^1$ with $\mathbb{Z}_N$ twisted boundary conditions to
the second order in a SUSY-breaking parameter~$\delta\epsilon$; this is the
quantity computed in~Ref.~\cite{Fujimori:2018kqp} by the bion calculus. First,
we find that the parameter~$\delta\epsilon$ receives renormalization and, after
this renormalization, the vacuum energy becomes ultraviolet (UV) finite. To the
next-to-leading order of the $1/N$ expansion, we find that the vacuum energy
is IR finite, as should be the case for a physical quantity. Finally, we find
that the vacuum energy normalized by the radius of the~$S^1$,
$RE(\delta\epsilon)$ behaves as inverse powers of~$\Lambda R$ for~$\Lambda R$
small, as shown in~Eqs.~\eqref{eq:(3.52)}--\eqref{eq:(3.57)}
and~Figs.~\ref{fig:2} and~\ref{fig:3}. Since $\Lambda$ is related to the
renormalized 't~Hooft coupling~$\lambda_R$ as~$\Lambda\sim e^{-2\pi/\lambda_R}$,
to the order of the $1/N$ expansion we work out, the vacuum energy is a purely
non-perturbative quantity and has \emph{no well-defined weak coupling expansion
in~$\lambda_R$\/}. This implies that one cannot even define the perturbative
expansion for this quantity computed in the $1/N$ expansion and cannot even
discuss the renormalon problem.\footnote{In Appendix~\ref{sec:A}, by taking
a particular limit~$R\to\infty$, we illustrate that the perturbative part of
the vacuum energy contains IR divergences, although when including the
non-perturbative part it becomes IR finite.} Therefore, although our $1/N$
calculation is robust, it does not give any clue to the issue. We do not yet
fully understand why the semi-classical calculation on the basis of the bion
cannot be observed in the $1/N$ expansion. Nevertheless, we believe that it is
worthwhile to report our $1/N$ calculation for future consideration because
our calculation itself is rather non-trivial.

\section{Two-dimensional SUSY $\mathbb{C}P^{N-1}$ model}
\label{sec:2}
\subsection{Action and boundary conditions}
Our spacetime is~$\mathbb{R}\times S^1$, and $-\infty<x<\infty$ denotes the
coordinate of~$\mathbb{R}$ and $0\leq y<2\pi R$ the coordinate of~$S^1$. The
Euclidean action of the 2D SUSY $\mathbb{C}P^{N-1}$ model in terms of the
homogeneous coordinate
variables~\cite{Cremmer:1978bh,DAdda:1978dle,Witten:1978bc} is, in the notation
of~Eq.~(2.24) of~Ref.~\cite{Ishikawa:2019tnw},
\begin{align}
   S&=\int d^2x\,\frac{N}{\lambda}\bigl[
   -f
   +\Bar{\sigma}\sigma
   +\Bar{z}^A(-D_\mu D_\mu+f)z^A
\notag\\
   &\qquad\qquad\qquad{}
   +\Bar{\chi}^A(\Slash{D}+\Bar{\sigma}P_++\sigma P_-)\chi^A
   +2\Bar{\chi}^Az^A\eta+2\Bar{\eta}\Bar{z}^A\chi^A
   \bigr]
\notag\\
   &\qquad{}
   -\int d^2x\,\frac{i\theta}{2\pi}\epsilon_{\mu\nu}\partial_\mu A_\nu.
\label{eq:(2.1)}
\end{align}
Here, and in what follows, it is understood that repeated indices are summed
over; the lower Greek indices, $\mu$, $\nu$, \dots, take the value $x$ or~$y$
and the uppercase Roman indices, $A$, $B$, \dots, run from~$1$ to~$N$.
$\lambda$ is the bare 't~Hooft coupling and $\theta$ is the theta
parameter.\footnote{The theta parameter~$\theta$ may be eliminated by the
anomalous chiral rotation
$\chi^A\to e^{i\alpha\gamma_5}\chi^A$,
$\Bar{\chi}^A\to\Bar{\chi}^Ae^{i\alpha\gamma_5}$,
$\eta\to e^{-i\alpha\gamma_5}\eta$,
$\Bar{\eta}\to\Bar{\eta}e^{-i\alpha\gamma_5}$,
and~$\sigma\to e^{2i\alpha}\sigma$.}
Also,
\begin{align}
   &D_\mu z^A\equiv(\partial_\mu+iA_\mu)z^A,\qquad
   \Slash{D}\chi^A
   \equiv\gamma_\mu(\partial_\mu+iA_\mu)\chi^A,
\notag\\
   &P_\pm\equiv\frac{1\pm\gamma_5}{2},\qquad
   \gamma_5\equiv-i\gamma_x\gamma_y,\qquad
   \gamma_x\equiv\begin{pmatrix}0&1\\1&0\\\end{pmatrix},\qquad
   \gamma_y\equiv\begin{pmatrix}0&-i\\i&0\\\end{pmatrix},
\label{eq:(2.2)}
\end{align}
and $\epsilon_{xy}=-\epsilon_{yx}=+1$.

For the fields with index~$A$ (we call them $N$-fields), we impose the
$\mathbb{Z}_N$~twisted boundary conditions along~$S^1$:
\begin{align}
   &z^A(x,y+2\pi R)=e^{2\pi im_AR}z^A(x,y),
\notag\\
   &\chi^A(x,y+2\pi R)=e^{2\pi im_AR}\chi^A(x,y),\qquad
   \Bar{\chi}^A(x,y+2\pi R)
   =e^{-2\pi im_AR}\Bar{\chi}^A(x,y),
\label{eq:(2.3)}
\end{align}
where the twist angle~$m_A$ in these expressions depends on the index~$A$ as
\begin{equation}
   \text{$m_A\equiv\frac{A}{NR}$ for~$A=1$, \dots, $N-1$},\qquad
   m_N\equiv0.
\label{eq:(2.4)}
\end{equation}
These twisted boundary conditions allow the fractional
instanton/anti-instanton, the constituent of the bion.

For the auxiliary fields, $f$, $\sigma$, $\Bar{\sigma}$, $A_\mu$, $\eta$,
and~$\Bar{\eta}$, on the other hand, we assume periodic boundary conditions
along~$S^1$.

For the calculation below, however, it turns out that an alternative form of
the action, obtained by
\begin{equation}
   f\to f+\Bar{\sigma}\sigma
\label{eq:(2.5)}
\end{equation}
from~Eq.~\eqref{eq:(2.1)}, that is,
\begin{align}
   S&=\int d^2x\,\frac{N}{\lambda}\bigl[
   -f+\Bar{z}^A(-D_\mu D_\mu+f+\Bar{\sigma}\sigma)z^A
\notag\\
   &\qquad\qquad\qquad{}
   +\Bar{\chi}^A(\Slash{D}+\Bar{\sigma}P_++\sigma P_-)\chi^A
   +2\Bar{\chi}^Az^A\eta+2\Bar{\eta}\Bar{z}^A\chi^A
   \bigr]
\notag\\
   &\qquad{}
   -\int d^2x\,\frac{i\theta}{2\pi}\epsilon_{\mu\nu}\partial_\mu A_\nu
\label{eq:(2.6)}
\end{align}
is more convenient. This is because renormalization with the action
in~Eq.~\eqref{eq:(2.1)} requires an infinite shift of the field~$f$ in addition
to the multiplicative renormalization of the 't~Hooft coupling
(Eq.~\eqref{eq:(2.10)} below), whereas the action in~Eq.~\eqref{eq:(2.6)} does
not require such a shift. This difference comes from the fact that
$\Bar{\sigma}\sigma$ in~Eq.~\eqref{eq:(2.5)} is a composite operator and UV
divergent. In fact, the action in~Eq.~\eqref{eq:(2.6)} can be obtained by the
dimensional reduction of a manifestly SUSY-invariant non-linear sigma model in
four dimensions~\cite{Lindstrom:1983rt}; we thus expect a simpler UV-divergent
structure. For this reason, we adopt the action in~Eq.~\eqref{eq:(2.6)} in the
present paper.

\subsection{Saddle point and propagators in the leading order of the $1/N$
expansion}
Now, since the action of~Eq.~\eqref{eq:(2.1)} (i.e.\ Eq.~(2.24)
of~Ref.~\cite{Ishikawa:2019tnw}) and the action of~Eq.~\eqref{eq:(2.6)} are
simply related by the change of variable in~Eq.~\eqref{eq:(2.5)}, we can borrow
the results in~Ref.~\cite{Ishikawa:2019tnw} in the leading order of the $1/N$
expansion.\footnote{With the twisted boundary conditions
of~Eq.~\eqref{eq:(2.3)}, as we will note in~Sect.~\ref{sec:3.1}, the effective
action arising from the Gaussian integration over $N$-fields is not simply
proportional to~$N$ but depends nontrivially on~$N$. Such a non-trivial
dependence on~$N$ in the Gaussian determinant is, however, exponentially
suppressed in the large-$N$ limit
of~Eq.~\eqref{eq:(1.1)}~\cite{Ishikawa:2019tnw} and can be neglected in
calculations in the $1/N$ expansion.}

First, setting
\begin{equation}
   A_\mu=A_{\mu0}+\delta A_\mu,\qquad
   f=f_0+\delta f,\qquad
   \sigma=\sigma_0+\delta\sigma,
\label{eq:(2.7)}
\end{equation}
where the subscript~$0$ indicates the value at the saddle point in the $1/N$
expansion and $\delta$~denotes the fluctuation, in the leading order of the
$1/N$ expansion in~Eq.~\eqref{eq:(1.1)} we have
\begin{equation}
   A_{\mu0}=A_{y0}\delta_{\mu y},\qquad f_0=0,\qquad
   \Bar{\sigma}_0\sigma_0=\Lambda^2,
\label{eq:(2.8)}
\end{equation}
where $\Lambda$ is the dynamical scale
\begin{equation}
   \Lambda=\mu e^{-2\pi/\lambda_R}
\label{eq:(2.9)}
\end{equation}
defined from the renormalized 't~Hooft coupling~$\lambda_R$ in the
``$\overline{\text{MS}}$~scheme,''
\begin{equation}
   \lambda=\left(\frac{e^{\gamma_E}\mu^2}{4\pi}\right)^\varepsilon\lambda_R
   \left(1+\frac{\lambda_R}{4\pi}\frac{1}{\varepsilon}\right)^{-1}.
\label{eq:(2.10)}
\end{equation}
Here, we have used dimensional regularization with the complex
dimension~$D=2-2\varepsilon$; $\mu$ is the renormalization scale.
In~Eq.~\eqref{eq:(2.8)}, the constant $A_{y0}$ is not determined from the
saddle point condition in the present supersymmetric theory and, for
$\mathbb{Z}_N$-invariant quantities such as the partition function and the
vacuum energy considered below, it should be integrated over with the
measure~\cite{Ishikawa:2019tnw}
\begin{equation}
   \int_0^1d(A_{y0}RN).
\label{eq:(2.11)}
\end{equation}

Next, we need the propagators among fluctuations of the auxiliary fields. To
obtain these, we add the gauge-fixing term
\begin{equation}
   S_{\text{gf}}
   =\frac{N}{4\pi}
   \int d^2x\,d^2x'\,
   \frac{1}{2}\partial_\mu\delta A_\mu(x)\partial_\nu\delta A_\nu(x')
   \int\frac{dp_x}{2\pi}\,\frac{1}{2\pi R}\sum_{p_y}
   e^{-ip(x-x')}\,\mathcal{L}(p)
\label{eq:(2.12)}
\end{equation}
and a local counter term
\begin{align}
   S_{\text{local}}
   &\equiv\frac{N}{4\pi}
   \int d^2x\,
   \left(-\frac{1}{2}\right)
   \left[\delta\sigma(x)-\delta\Bar{\sigma}(x)\right]^2
\label{eq:(2.13)}
\end{align}
to the action in~Eq.~\eqref{eq:(2.6)}~\cite{Ishikawa:2019tnw}. Then, in the
leading order of the $1/N$ expansion, we have
\begin{align}
   &\left\langle\delta A_\mu(x)\delta A_\nu(x')
   \right\rangle
\notag\\
   &=\frac{4\pi}{N}\int\frac{dp_x}{2\pi}\,\frac{1}{2\pi R}\sum_{p_y}
   e^{ip(x-x')}\,
   \frac{\mathcal{L}(p)}{\mathcal{D}(p)}
   \left\{\delta_{\mu\nu}
   +4\left[\Lambda^2+\frac{\Bar{p}_y^2}{p^2}
   \frac{\mathcal{K}(p)^2}{\mathcal{L}(p)^2}
   \right]
   \frac{p_\mu p_\nu}{(p^2)^2}
   \right\},
\notag\\
   &\left\langle\delta A_\mu(x)\delta R(x')
   \right\rangle
   =\left\langle\delta R(x)\delta A_\mu(x')
   \right\rangle=0,
\notag\\
   &\left\langle\delta A_\mu(x)\delta I(x')
   \right\rangle
   =-\left\langle\delta I(x)\delta A_\mu(x')
   \right\rangle
   =\frac{4\pi}{N}\int\frac{dp_x}{2\pi}\,\frac{1}{2\pi R}\sum_{p_y}
   e^{ip(x-x')}\,\frac{\mathcal{L}(p)}{\mathcal{D}(p)}
   \frac{2\Lambda^2\Bar{p}_\mu}{p^2},
\notag\\
   &\left\langle\delta A_\mu(x)\delta f(x')
   \right\rangle
   =\left\langle\delta f(x)\delta A_\mu(x')
   \right\rangle
   =\frac{4\pi}{N}\int\frac{dp_x}{2\pi}\,\frac{1}{2\pi R}\sum_{p_y}
   e^{ip(x-x')}\,\frac{\mathcal{K}(p)}{\mathcal{D}(p)}
   \frac{-2\Bar{p}_\mu\Bar{p}_y}{p^2},
\notag\\
   &\left\langle\delta R(x)\delta R(x')
   \right\rangle
   =\frac{4\pi}{N}\int\frac{dp_x}{2\pi}\,\frac{1}{2\pi R}\sum_{p_y}
   e^{ip(x-x')}\,\frac{\mathcal{L}(p)}{\mathcal{D}(p)}\Lambda^2,
\notag\\
   &\left\langle\delta R(x)\delta I(x')
   \right\rangle
   =-\left\langle\delta I(x)\delta R(x')
   \right\rangle
   =\frac{4\pi}{N}\int\frac{dp_x}{2\pi}\,\frac{1}{2\pi R}\sum_{p_y}
   e^{ip(x-x')}\,\frac{\mathcal{K}(p)}{\mathcal{D}(p)}
   \frac{-2\Lambda^2\Bar{p}_y}{p^2},
\notag\\
   &\left\langle\delta R(x)\delta f(x')
   \right\rangle
   =\left\langle\delta f(x)\delta R(x')
   \right\rangle
   =\frac{4\pi}{N}\int\frac{dp_x}{2\pi}\,\frac{1}{2\pi R}\sum_{p_y}
   e^{ip(x-x')}\,\frac{\mathcal{L}(p)}{\mathcal{D}(p)}(-2\Lambda^2),
\notag\\
   &\left\langle\delta I(x)\delta I(x')
   \right\rangle
   =\frac{4\pi}{N}\int\frac{dp_x}{2\pi}\,\frac{1}{2\pi R}\sum_{p_y}
   e^{ip(x-x')}\,\frac{\mathcal{L}(p)}{\mathcal{D}(p)}\Lambda^2,
\notag\\
   &\left\langle\delta I(x)\delta f(x')
   \right\rangle
   =-\left\langle\delta f(x)\delta I(x')
   \right\rangle
   =0,
\notag\\
   &\left\langle\delta f(x)\delta f(x')
   \right\rangle
   =\frac{4\pi}{N}\int\frac{dp_x}{2\pi}\,\frac{1}{2\pi R}\sum_{p_y}
   e^{ip(x-x')}\,\frac{\mathcal{L}(p)}{\mathcal{D}(p)}(-p^2),
\notag\\
   &\left\langle\eta(x)\Bar{\eta}(x')\right\rangle
\notag\\
   &=\frac{4\pi}{N}\int\frac{dp_x}{2\pi}\,\frac{1}{2\pi R}\sum_{p_y}
   e^{ip(x-x')}\,
   \frac{
   (i\Slash{p}+2\Bar{\sigma}_0P_++2\sigma_0P_-)\mathcal{L}(p)
   +2i\Slash{\Bar{p}}\Bar{p}_y/p^2\mathcal{K}(p)}
   {\mathcal{D}(p)}\left(-\frac{1}{2}\right),
\label{eq:(2.14)}
\end{align}
where the Kaluza--Klein (KK) momentum along~$S^1$, $p_y$, takes discrete values
$p_y=n/R$ with~$n\in\mathbb{Z}$. We have also introduced the notations
\begin{equation}
   \Bar{p}_\mu\equiv\epsilon_{\nu\mu}p_\nu
\label{eq:(2.15)}
\end{equation}
and
\begin{equation}
   \delta R(x)
   \equiv\frac{1}{2}\left[
   \Bar{\sigma}_0\delta\sigma(x)
   +\sigma_0\delta\Bar{\sigma}(x)
   \right],\qquad
   \delta I(x)
   \equiv\frac{1}{2i}\left[
   \Bar{\sigma}_0\delta\sigma(x)
   -\sigma_0\delta\Bar{\sigma}(x)
   \right].
\label{eq:(2.16)}
\end{equation}
From the above results, we also have
\begin{equation}
   \left\langle\delta\sigma(x)\delta\Bar{\sigma}(x')
   \right\rangle
   =\frac{4\pi}{N}\int\frac{dp_x}{2\pi}\,\frac{1}{2\pi R}\sum_{p_y}
   e^{ip(x-x')}\,\frac{1}{\mathcal{D}(p)}
   \left[2\mathcal{L}(p)
   +4i\frac{\Bar{p}_y}{p^2}\mathcal{K}(p)
   \right].
\label{eq:(2.17)}
\end{equation}

Various functions used in the above expressions are defined by
\begin{align}
   \mathcal{L}(p)&\equiv\mathcal{L}_\infty(p)+\Hat{\mathcal{L}}(p),
\notag\\
   \mathcal{L}_\infty(p)
   &\equiv\frac{2}{\sqrt{p^2(p^2+4\Lambda^2)}}
   \ln\left(\frac{\sqrt{p^2+4\Lambda^2}+\sqrt{p^2}}
   {\sqrt{p^2+4\Lambda^2}-\sqrt{p^2}}\right),
\notag\\
   \Hat{\mathcal{L}}(p)&\equiv
   \int_0^1dx\,
   \sum_{m\neq0}e^{-iA_{y0}2\pi RNm}e^{ixp_y2\pi RNm}
\notag\\
   &\qquad\qquad\qquad{}
   \times\frac{2\pi RN|m|}{\sqrt{\Lambda^2+x(1-x)p^2}}
   K_1(\sqrt{\Lambda^2+x(1-x)p^2}2\pi RN|m|),
\notag\\
   \mathcal{K}(p)&\equiv
   i\int_0^1dx\,\sum_{m\neq0}e^{-iA_{y0}2\pi RNm}e^{ixp_y2\pi RNm}2\pi RNm
   K_0(\sqrt{\Lambda^2+x(1-x)p^2}2\pi RN|m|),
\notag\\
   \mathcal{D}(p)
   &\equiv(p^2+4\Lambda^2)\mathcal{L}(p)^2+4\frac{\Bar{p}_y^2}{p^2}
   \mathcal{K}(p)^2,
\label{eq:(2.18)}
\end{align}
where $K_\nu(z)$ denotes the modified Bessel function of the second kind. For
later calculations, it is important to note the properties
\begin{equation}
   \mathcal{L}(p)=\mathcal{L}(-p),\qquad
   \mathcal{K}(p)=\mathcal{K}(-p).
\label{eq:(2.19)}
\end{equation}
These can be shown by the change of the Feynman parameter, $x\to1-x$, noting
that $p_y\in\mathbb{Z}/R$.

Going back to the action in~Eq.~$S$~\eqref{eq:(2.6)}, with the saddle point
values in~Eq.~\eqref{eq:(2.8)}, the propagators of the $N$-fields in the
leading order of the $1/N$ expansion are given by
\begin{align}
   \left\langle
   z^A(x)\Bar{z}^B(x')
   \right\rangle
   &=\delta^{AB}\frac{\lambda}{N}
   \int\frac{dp_x}{2\pi}\,\frac{1}{2\pi R}\sum_{p_y}
   e^{ip_x(x-x')}e^{i(p_y+m_A)(y-y')}\,
\notag\\
   &\qquad\qquad\qquad\qquad\qquad{}
   \times\left[p_x^2+(p_y+A_{y0}+m_A)^2+\Lambda^2\right]^{-1},
\notag\\
   \left\langle
   \chi^A(x)\Bar{\chi}^B(x')
   \right\rangle
   &=\delta^{AB}\frac{\lambda}{N}
   \int\frac{dp_x}{2\pi}\,\frac{1}{2\pi R}\sum_{p_y}
   e^{ip_x(x-x')}e^{i(p_y+m_A)(y-y')}\,
\notag\\
   &\qquad\qquad\qquad{}
   \times\left[i\gamma_xp_x+i\gamma_y(p_y+A_{y0}+m_A)
   +\Bar{\sigma}_0P_++\sigma_0P_-\right]^{-1}.
\label{eq:(2.20)}
\end{align}
To obtain these, we noted the twisted boundary conditions
of~Eq.~\eqref{eq:(2.3)}.

\section{Computation of the vacuum energy}
\label{sec:3}
\subsection{General strategy}
\label{sec:3.1}
Our objective in this paper is to compute the vacuum energy of the present
system as a power series of the coefficient~$\delta\epsilon$ of a SUSY-breaking
term---the quantity computed
in~Ref.~\cite{Fujimori:2018kqp}:
\begin{equation}
   E(\delta\epsilon)
   =E^{(0)}+E^{(1)}\delta\epsilon+E^{(2)}\delta\epsilon^2+\dotsb.
\label{eq:(3.1)}
\end{equation}
Here, the supersymmetry breaking term introduced
in~Ref.~\cite{Fujimori:2018kqp} is
\begin{equation}
   \delta S
   \equiv\int d^2x\,\frac{\delta\epsilon}{\pi R}\sum_{A=1}^Nm_A
   \left(\Bar{z}^Az^A-\frac{1}{N}\right).
\label{eq:(3.2)}
\end{equation}
Note that this depends on the twist angles in~Eq.~\eqref{eq:(2.4)}. A quick way
to incorporate the effect of~Eq.~\eqref{eq:(3.2)} is to regard $\delta S$ as a
mass term of the $z^A$-field, as
\begin{align}
   S+\delta S&=\int d^2x\,\frac{N}{\lambda}
   \Bar{z}^A\left(-\partial_\mu\partial_\mu+\Lambda^2
   +\delta_A\right)z^A
   +\dotsb,
\label{eq:(3.3)}
\end{align}
where
\begin{equation}
   \delta_A\equiv\frac{\lambda\delta\epsilon}{\pi RN}m_A.
\label{eq:(3.4)}
\end{equation}
With this modification, the vacuum energy is given by
\begin{align}
   -\int dx\,E(\delta\epsilon)
   &=\int d^2x\,\frac{1}{\lambda}\sum_A\delta_A
   -\sum_A\ln\Det(-\partial_\mu\partial_\mu+\Lambda^2+\delta_A)
\notag\\
   &\qquad{}
   +(\text{connected vacuum bubble diagrams}).
\label{eq:(3.5)}
\end{align}
Here, the vacuum bubble diagrams, which start from two-loop order, are
computed by using the modified $z^A$-propagator
\begin{align}
   &\left\langle
   z^A(x)\Bar{z}^B(x')
   \right\rangle
\notag\\
   &=\delta^{AB}\frac{\lambda}{N}
   \int\frac{dp_x}{2\pi}\,\frac{1}{2\pi R}\sum_{p_y}
   e^{ip_x(x-x')}e^{i(p_y+m_A)(y-y')}\,
   \left[p_x^2+(p_y+A_{y0}+m_A)^2+\Lambda^2+\delta_A\right]^{-1}
\label{eq:(3.6)}
\end{align}
instead of the one in~Eq.~\eqref{eq:(2.20)}. Then, by expanding
Eq.~\eqref{eq:(3.5)} with respect to~$\delta_A$, we have the series expansion
in~Eq.~\eqref{eq:(3.1)}. In the following calculations, we set $E^{(0)}=0$
assuming that the bare vacuum energy at~$\delta\epsilon=0$ is chosen so that
the system is supersymmetric for~$\delta\epsilon=0$. This amounts to computing
the difference $E(\delta\epsilon)-E(\delta\epsilon=0)$.

If all the $N$-fields obey the same boundary conditions along~$S^1$, all $z^A$
(or $\chi^A$ and~$\Bar{\chi}^A$) contribute equally and the order of the loop
expansion with the use of the auxiliary fields and the order of the $1/N$
expansion would coincide~\cite{Coleman:1985rnk}. With the twisted boundary
conditions in~Eq.~\eqref{eq:(2.3)}, however, not all $N$-fields contribute
equally. The SUSY-breaking term in~Eq.~\eqref{eq:(3.2)} also treats each of
$N$-fields differently. For these reasons, in the present system the order of
the loop expansion and that of the $1/N$ expansion do not necessarily coincide;
we have to distinguish both expansions. For instance, although the one-loop
Gaussian determinant in~Eq.~\eqref{eq:(3.5)} gives rise to the contribution
of~$O(1/N)$, it also contains terms of subleading orders, $O(1/N^2)$
and~$O(1/N^3)$ (see~Eq.~\eqref{eq:(3.49)}, for instance).

\subsection{One-loop Gaussian determinant}
Let us start with the one-loop Gaussian determinant in~Eq.~\eqref{eq:(3.5)}.
We first note that
\begin{align}
   &-\sum_A\ln\Det(-\partial_\mu\partial_\mu+\Lambda^2+\delta_A)
\notag\\
   &=-\sum_A\int d^2x\,
   \int\frac{dp_x}{2\pi}\frac{1}{2\pi R}\sum_{p_y}
   \ln\left[p_x^2+(p_y+m_A+A_{y0})^2+\Lambda^2
   +\delta_A\right]
\notag\\
   &=-\int d^2x\,\sum_A\sum_{n=-\infty}^\infty
   \int\frac{d^2p}{(2\pi)^2}\,e^{i(p_y-m_A-A_{y0})2\pi Rn}
   \ln(p^2+\Lambda^2+\delta_A),
\label{eq:(3.7)}
\end{align}
where we have used the identity
\begin{equation}
   \frac{1}{2\pi R}\sum_{n=-\infty}^\infty F(n/R)
   =\sum_{n=-\infty}^\infty\int\frac{dp_y}{2\pi}\,e^{ip_y2\pi Rn}F(p_y).
\label{eq:(3.8)}
\end{equation}
Hence, subtracting the logarithm of the Gaussian determinant
at~$\delta\epsilon=0$, we have
\begin{align}
   &-\sum_A\ln\Det\left(
   \frac{-\partial_\mu\partial_\mu+\Lambda^2+\delta_A}
   {-\partial_\mu\partial_\mu+\Lambda^2}\right)
\notag\\
   &=-\int d^2x\,\sum_A\sum_{n=-\infty}^\infty
   \int\frac{d^2p}{(2\pi)^2}\,e^{i(p_y-m_A-A_{y0})2\pi Rn}
   \ln\left(
   \frac{p^2+\Lambda^2+\delta_A}
   {p^2+\Lambda^2}\right).
\label{eq:(3.9)}
\end{align}
In this expression, since the $n\neq0$ terms are Fourier transforms, only the
$n=0$ term is UV divergent. Under the dimensional regularization
with~$D=2-2\varepsilon$, the momentum integration yields
\begin{align}
   &-\sum_A\ln\Det\left(
   \frac{-\partial_\mu\partial_\mu+\Lambda^2+\delta_A}
   {-\partial_\mu\partial_\mu+\Lambda^2}\right)
\notag\\
   &=-\int d^2x\,\frac{1}{4\pi}\left[\frac{1}{\varepsilon}
   -\ln\left(\frac{e^{\gamma_E}\Lambda^2}{4\pi}\right)\right]\sum_A
   \delta_A
\notag\\
   &\qquad{}
   -\int d^2x\,\sum_A
   \frac{1}{4\pi}
   \left[\delta_A
   -(\Lambda^2+\delta_A)
   \ln\left(1+\frac{\delta_A}{\Lambda^2}\right)
   \right]
\notag\\
   &\qquad{}
   -\int d^2x\,\sum_A
   \sum_{n\neq0}e^{-i(m_A+A_{y0})2\pi Rn}
\notag\\
   &\qquad\qquad{}
   \times\frac{1}{4\pi}
   (-4)\frac{1}{2\pi R|n|}
   \left[
   \sqrt{\Lambda^2+\delta_A}
   K_1(\sqrt{\Lambda^2+\delta_A}2\pi R|n|)
   -\Lambda K_1(\Lambda2\pi R|n|)
   \right].
\label{eq:(3.10)}
\end{align}
Since Eqs.~\eqref{eq:(2.9)} and~\eqref{eq:(2.10)} imply that
\begin{equation}
   \frac{1}{4\pi}
   \left[\frac{1}{\varepsilon}
   -\ln\left(\frac{e^{\gamma_E}\Lambda^2}{4\pi}\right)\right]
   =\frac{1}{\lambda},
\label{eq:(3.11)}
\end{equation}
we see that the first term on the right-hand side of~Eq.~\eqref{eq:(3.10)} is
precisely canceled by the first term on the right-hand side
of~Eq.~\eqref{eq:(3.5)}.

In this way, from~Eq.~\eqref{eq:(3.5)} we have
\begin{align}
   &\left.E(\delta\epsilon)\right|_{\text{1-loop}}
\notag\\
   &=2\pi R\sum_A
   \frac{1}{4\pi}
   \left[\delta_A
   -(\Lambda^2+\delta_A)
   \ln\left(1+\frac{\delta_A}{\Lambda^2}\right)
   \right]
\notag\\
   &\qquad{}
   +2\pi R\sum_A
   \sum_{n\neq0}e^{-i(m_A+A_{y0})2\pi Rn}
\notag\\
   &\qquad\qquad{}
   \times\frac{1}{4\pi}
   (-4)\frac{1}{2\pi R|n|}
   \left[
   \sqrt{\Lambda^2+\delta_A}
   K_1(\sqrt{\Lambda^2+\delta_A}2\pi R|n|)
   -\Lambda K_1(\Lambda2\pi R|n|)
   \right].
\label{eq:(3.12)}
\end{align}
\subsection{Two-loop vacuum bubble diagrams}
\begin{figure}
\centering
\begin{subfigure}{0.12\columnwidth}
\centering
\includegraphics[width=\columnwidth]{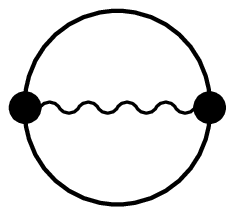}
\caption{}
\label{fig:1(a)}
\end{subfigure}
\hspace{5mm}
\begin{subfigure}{0.12\columnwidth}
\centering
\includegraphics[width=\columnwidth]{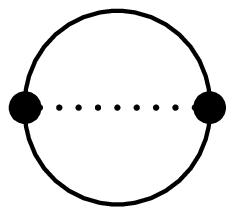}
\caption{}
\label{fig:1(b)}
\end{subfigure}
\hspace{5mm}
\begin{subfigure}{0.12\columnwidth}
\centering
\includegraphics[width=\columnwidth]{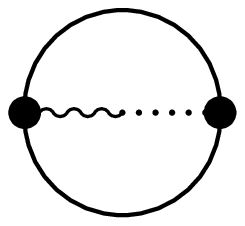}
\caption{}
\label{fig:1(c)}
\end{subfigure}
\hspace{5mm}
\begin{subfigure}{0.12\columnwidth}
\centering
\includegraphics[width=\columnwidth]{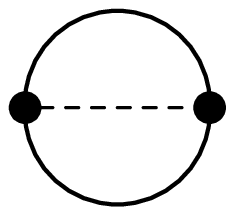}
\caption{}
\label{fig:1(d)}
\end{subfigure}
\hspace{5mm}
\begin{subfigure}{0.12\columnwidth}
\centering
\includegraphics[width=\columnwidth]{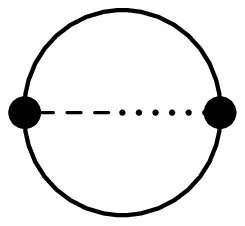}
\caption{}
\label{fig:1(e)}
\end{subfigure}
\hspace{5mm}
\begin{subfigure}{0.12\columnwidth}
\centering
\includegraphics[width=\columnwidth]{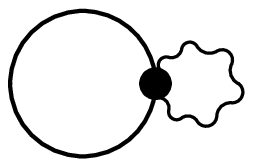}
\caption{}
\label{fig:1(f)}
\end{subfigure}
\hspace{5mm}
\begin{subfigure}{0.12\columnwidth}
\centering
\includegraphics[width=\columnwidth]{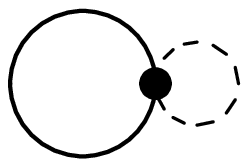}
\caption{}
\label{fig:1(g)}
\end{subfigure}
\hspace{5mm}
\begin{subfigure}{0.12\columnwidth}
\centering
\includegraphics[width=\columnwidth]{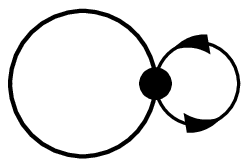}
\caption{}
\label{fig:1(h)}
\end{subfigure}
\caption{Two-loop vacuum bubble diagrams that contribute
to~$E(\delta\epsilon)|_{\text{2-loop}}$ in~Eq.~\eqref{eq:(3.13)}. The solid line
denotes the $z^A$-propagator of~Eq.~\eqref{eq:(3.6)}. The wavy line denotes the
$\delta A_\mu$-propagator, the dotted line the $\delta f$-propagator, the
broken line the $\delta\sigma$-propagator, and the arrowed solid line the
$\eta$-propagator in~Eqs.~\eqref{eq:(2.17)} and~\eqref{eq:(2.14)}.}
\label{fig:1}
\end{figure}

Next, we work out the vacuum bubble diagrams in the two-loop level; they are
depicted in~Fig.~\ref{fig:1}.
By using the propagators in~Eqs.~\eqref{eq:(2.14)}, \eqref{eq:(2.17)},
\eqref{eq:(2.20)}, and~\eqref{eq:(3.6)}, and interaction vertices
in~Eq.~\eqref{eq:(2.6)}, from~Eq.~\eqref{eq:(3.5)} we have
\begin{align}
   &\left.E(\delta\epsilon)\right|_{\text{2-loop}}
\notag\\
   &=-2\pi R\frac{4\pi}{N}\sum_A\sum_{n=-\infty}^\infty
   \int\frac{d^2p}{(2\pi)^2}\,e^{i(p_y-A_{y0}-m_A)2\pi Rn}
   \frac{1}{p^2+\Lambda^2+\delta_A}
\notag\\
   &\qquad{}
   \times
   \Biggl[
   \int\frac{d\ell_x}{2\pi}\frac{1}{2\pi R}\sum_{\ell_y}
   \frac{1}{(p-\ell)^2+\Lambda^2+\delta_A}
\notag\\
   &\qquad\qquad{}
   \times\Biggl(
   \frac{1}{2}(2p_\mu-\ell_\mu)(2p_\nu-\ell_\nu)
   \frac{\mathcal{L}(\ell)}{\mathcal{D}(\ell)}
   \left\{
   \delta_{\mu\nu}+4\left[
   \Lambda^2+\frac{\Bar{\ell}_y^2}{\ell^2}
   \frac{\mathcal{K}(\ell)^2}{\mathcal{L}(\ell)^2}
   \right]\frac{\ell_\mu\ell_\nu}{(\ell^2)^2}
   \right\}
\tag{Fig.~\ref{fig:1(a)}}\\
   &\qquad\qquad\qquad{}
   -\frac{1}{2}\frac{\mathcal{L}(\ell)}{\mathcal{D}(\ell)}\ell^2
\tag{Fig.~\ref{fig:1(b)}}\\
   &\qquad\qquad\qquad{}
   -4p_\mu\frac{\mathcal{K}(\ell)}{\mathcal{D}(\ell)}
   \frac{\Bar{\ell}_\mu\Bar{\ell}_y}{\ell^2}
\tag{Fig.~\ref{fig:1(c)}}\\
   &\qquad\qquad\qquad{}
   +2\frac{\mathcal{L}(\ell)}{\mathcal{D}(\ell)}\Lambda^2
\tag{Fig.~\ref{fig:1(d)}}\\
   &\qquad\qquad\qquad{}
   -4\frac{\mathcal{L}(\ell)}{\mathcal{D}(\ell)}\Lambda^2
   \Biggr)
\tag{Fig.~\ref{fig:1(e)}}\\
   &\qquad\qquad{}
   +\int\frac{d\ell_x}{2\pi}\frac{1}{2\pi R}\sum_{\ell_y}
\notag\\
   &\qquad\qquad\qquad{}
   \times\Biggl(-
   \frac{\mathcal{L}(\ell)}{\mathcal{D}(\ell)}
   \left\{
   2+4\left[
   \Lambda^2+\frac{\Bar{\ell}_y^2}{\ell^2}
   \frac{\mathcal{K}(\ell)^2}{\mathcal{L}(\ell)^2}
   \right]\frac{1}{\ell^2}
   \right\}
\tag{Fig.~\ref{fig:1(f)}}\\
   &\qquad\qquad\qquad\qquad{}
   -2\frac{\mathcal{L}(\ell)}{\mathcal{D}(\ell)}
\tag{Fig.~\ref{fig:1(g)}}\\
   &\qquad\qquad\qquad\qquad{}
   +\frac{1}{(p-\ell)^2+\Lambda^2}
   4\left\{
   [-(p-\ell)\cdot\ell+2\Lambda^2]
   \frac{\mathcal{L}(\ell)}{\mathcal{D}(\ell)}
   +2p_\mu\frac{\Bar{\ell}_\mu\Bar{\ell}_y}{\ell^2}
   \frac{\mathcal{K}(\ell)}{\mathcal{D}(\ell)}
   \right\}
   \Biggr)\Biggr]
\tag{Fig.~\ref{fig:1(h)}}\\
   &\qquad{}
   -(\text{terms with $\delta\epsilon=0$}),
\label{eq:(3.13)}
\end{align}
where the contributions of each diagram in~Fig.~\ref{fig:1} are separately
indicated by the equation numbers. The total sum is
\begin{align}
   &\left.E(\delta\epsilon)\right|_{\text{2-loop}}
\notag\\
   &=-2\pi R\frac{4\pi}{N}\sum_A\sum_{n=-\infty}^\infty
   \int\frac{d^2p}{(2\pi)^2}\,e^{i(p_y-A_{y0}-m_A)2\pi Rn}
   \frac{1}{p^2+\Lambda^2+\delta_A}
\notag\\
   &\qquad{}
   \times
   \Biggl(
   \int\frac{d\ell_x}{2\pi}\frac{1}{2\pi R}\sum_{\ell_y}
   \frac{1}{(p-\ell)^2+\Lambda^2+\delta_A}
\notag\\
   &\qquad\qquad{}
   \times\biggl\{
   \frac{\mathcal{L}(\ell)}{\mathcal{D}(\ell)}
   \left[
   2p^2-2p\cdot\ell-8\Lambda^2\frac{p\cdot\ell}{\ell^2}
   +8\Lambda^2\frac{(p\cdot\ell)^2}{(\ell^2)^2}
   \right]
\notag\\
   &\qquad\qquad\qquad{}
   +\frac{\mathcal{K}(\ell)^2}{\mathcal{D}(\ell)\mathcal{L}(\ell)}
   \frac{\Bar{\ell}_y^2}{\ell^2}
   \left[
   2-8\frac{p\cdot\ell}{\ell^2}+8\frac{(p\cdot\ell)^2}{(\ell^2)^2}
   \right]
   +\frac{\mathcal{K}(\ell)}{\mathcal{D}(\ell)}(-4)
   \frac{p\cdot\Bar{\ell}\,\Bar{\ell}_y}{\ell^2}\biggr\}
\notag\\
   &\qquad\qquad{}
   +\int\frac{d\ell_x}{2\pi}\frac{1}{2\pi R}\sum_{\ell_y}
   \frac{1}{(p-\ell)^2+\Lambda^2}
\notag\\
   &\qquad\qquad\qquad{}
   \times\biggl\{
   \frac{\mathcal{L}(\ell)}{\mathcal{D}(\ell)}
   \left[-4p^2+4p\cdot\ell
   +8\Lambda^2\frac{p\cdot\ell}{\ell^2}
   -4\Lambda^2\frac{p^2}{\ell^2}
   -4\Lambda^4\frac{1}{\ell^2}
   \right]
\notag\\
   &\qquad\qquad\qquad\qquad{}
   +\frac{\mathcal{K}(\ell)^2}{\mathcal{D}(\ell)\mathcal{L}(\ell)}
   \frac{\Bar{\ell}_y^2}{\ell^2}
   \left[-4+8\frac{p\cdot\ell}{\ell^2}-4\frac{p^2}{\ell^2}
   -4\Lambda^2\frac{1}{\ell^2}\right]
   +\frac{\mathcal{K}(\ell)}{\mathcal{D}(\ell)}(8)
   \frac{p\cdot\Bar{\ell}\,\Bar{\ell}_y}{\ell^2}
   \biggr\}\Biggr)
\notag\\
   &\qquad{}
   -(\text{term with $\delta\epsilon=0$}).
\label{eq:(3.14)}
\end{align}
To examine the renormalizability of this expression, we first note that this
can be written as
\begin{align}
   &\left.E(\delta\epsilon)\right|_{\text{2-loop}}
\notag\\
   &=-2\pi R\frac{4\pi}{N}\sum_A
   \left.\left\{
   \left(e^{\delta_A\partial_\xi}e^{\delta_A\partial_\eta}-1\right)
   I(\xi,\eta)
   +\left(e^{\delta_A\partial_\xi}-1\right)
   \left[-2I(\xi,0)+J(\xi)\right]
   \right\}\right|_{\xi=\eta=0},
\label{eq:(3.15)}
\end{align}
where
\begin{align}
   I(\xi,\eta)
   &\equiv\int\frac{d\ell_x}{2\pi}\frac{1}{2\pi R}\sum_{\ell_y}
   \sum_{n=-\infty}^\infty\int\frac{d^2p}{(2\pi)^2}\,e^{i(p_y-A_{y0}-m_A)2\pi Rn}
   \frac{1}{p^2+\Lambda^2+\xi}
   \frac{1}{(p-\ell)^2+\Lambda^2+\eta}
\notag\\
   &\qquad{}
   \times\biggl\{
   \frac{\mathcal{L}(\ell)}{\mathcal{D}(\ell)}
   \left[
   2p^2-2p\cdot\ell-8\Lambda^2\frac{p\cdot\ell}{\ell^2}
   +8\Lambda^2\frac{(p\cdot\ell)^2}{(\ell^2)^2}
   \right]
\notag\\
   &\qquad\qquad{}
   +\frac{\mathcal{K}(\ell)^2}{\mathcal{D}(\ell)\mathcal{L}(\ell)}
   \frac{\Bar{\ell}_y^2}{\ell^2}
   \left[
   2-8\frac{p\cdot\ell}{\ell^2}+8\frac{(p\cdot\ell)^2}{(\ell^2)^2}
   \right]
   +\frac{\mathcal{K}(\ell)}{\mathcal{D}(\ell)}(-4)
   \frac{p\cdot\Bar{\ell}\,\Bar{\ell}_y}{\ell^2}\biggr\},
\label{eq:(3.16)}
\end{align}
and
\begin{align}
   J(\xi)
   &\equiv\int\frac{d\ell_x}{2\pi}\frac{1}{2\pi R}\sum_{\ell_y}
   \sum_{n=-\infty}^\infty\int\frac{d^2p}{(2\pi)^2}\,e^{i(p_y-A_{y0}-m_A)2\pi Rn}
   \frac{1}{p^2+\Lambda^2+\xi}
   \frac{1}{(p-\ell)^2+\Lambda^2}
\notag\\
   &\qquad{}
   \times
   \left[
   \frac{\mathcal{L}(\ell)}{\mathcal{D}(\ell)}\Lambda^2
   +\frac{\mathcal{K}(\ell)^2}{\mathcal{D}(\ell)\mathcal{L}(\ell)}
   \frac{\Bar{\ell}_y^2}{\ell^2}
   \right]
   \left[
   -8\frac{p\cdot\ell}{\ell^2}
   -4\frac{p^2}{\ell^2}
   -4\Lambda^2\frac{1}{\ell^2}
   +16\frac{(p\cdot\ell)^2}{(\ell^2)^2}
   \right].
\label{eq:(3.17)}
\end{align}

From Eq.~\eqref{eq:(2.18)}, we see that, for $|\ell|\to\infty$,
$\Hat{\mathcal{L}}(p)$ and~$\mathcal{K}(p)$ are exponentially small because
of the Bessel functions, and thus
\begin{equation}
   \mathcal{L}(\ell)
   \stackrel{|\ell|\to\infty}{\to}\frac{2}{\ell^2}\ln(\ell^2/\Lambda^2),\qquad
   \mathcal{D}(\ell)
   \stackrel{|\ell|\to\infty}{\to}\ell^2\mathcal{L}(\ell)^2.
\label{eq:(3.18)}
\end{equation}
From these, we see that, in~$I(\xi,\eta)$ of~Eq.~\eqref{eq:(3.16)}, the
integration over~$\ell$ as well as the integration over~$p$ are logarithmically
UV divergent. In $J(\xi)$ of~Eq.~\eqref{eq:(3.17)}, the integration over~$p$ is
logarithmically UV divergent but the integration over~$\ell$ is UV convergent.
Assuming (say) the dimensional regularization, the change of integration
variables $(p,\ell)\to(p-\ell,-\ell)$ in~$I(\xi,\eta)$, Eq.~\eqref{eq:(3.16)},
shows that
\begin{equation}
   I(\xi,\eta)=I(\eta,\xi).
\label{eq:(3.19)}
\end{equation}

Now, in~Eq.~\eqref{eq:(3.15)}, using the identity
\begin{equation}
   e^{\delta_A\partial_\xi}e^{\delta_A\partial_\eta}-1
   =\left(e^{\delta_A\partial_\xi}-1\right)
   \left(e^{\delta_A\partial_\eta}-1\right)
   +e^{\delta_A\partial_\xi}
   +e^{\delta_A\partial_\eta}-2
\label{eq:(3.20)}
\end{equation}
and noting the property in~Eq.~\eqref{eq:(3.19)}, we have the following very
convenient representation:
\begin{align}
   &\left.E(\delta\epsilon)\right|_{\text{2-loop}}
\notag\\
   &=-2\pi R\frac{4\pi}{N}\sum_A
   \left.\left[
   \left(e^{\delta_A\partial_\xi}-1\right)
   \left(e^{\delta_A\partial_\eta}-1\right)
   I(\xi,\eta)
   +\left(e^{\delta_A\partial_\xi}-1\right)
   J(\xi)
   \right]\right|_{\xi=\eta=0}.
\label{eq:(3.21)}
\end{align}
This shows that $E(\delta\epsilon)|_{\text{2-loop}}$ is UV finite \emph{provided
that the parameter~$\delta_A$ is UV finite}. That is, the operator
$e^{\delta_A\partial_\xi}-1$ acting on~$J(\xi)$ increases the power
of~$p^2+\Lambda^2$ in the denominator in~Eq.~\eqref{eq:(3.17)} and makes the
$p$~integration UV finite. Similarly, the operator
$(e^{\delta_A\partial_\xi}-1)(e^{\delta_A\partial_\eta}-1)$ acting on~$I(\xi,\eta)$
increases the power of~$(p^2+\Lambda^2)[(p-\ell)^2+\Lambda^2]$ in the
denominator of~Eq.~\eqref{eq:(3.16)} and makes the integrations over $p$
and~$\ell$ UV convergent.

\subsection{Renormalizability to the two-loop order}
So far, we have observed that, from~Eq.~\eqref{eq:(3.12)},
\begin{align}
   &\left.E(\delta\epsilon)\right|_{\text{1-loop}}
\notag\\
   &=2\pi R\sum_A
   \frac{1}{4\pi}
   \left[\delta_A
   -(\Lambda^2+\delta_A)
   \ln\left(1+\frac{\delta_A}{\Lambda^2}\right)
   \right]
\notag\\
   &\qquad{}
   +2\pi R\sum_A
   \sum_{n\neq0}e^{-i(m_A+A_{y0})2\pi Rn}
\notag\\
   &\qquad\qquad{}
   \times\frac{1}{4\pi}
   (-4)\frac{1}{2\pi R|n|}
   \left[
   \sqrt{\Lambda^2+\delta_A}
   K_1(\sqrt{\Lambda^2+\delta_A}2\pi R|n|)
   -\Lambda K_1(\Lambda2\pi R|n|)
   \right],
\label{eq:(3.22)}
\end{align}
and, from~Eq.~\eqref{eq:(3.21)},
\begin{equation}
   \left.E(\delta\epsilon)\right|_{\text{2-loop}}
   =-2\pi R\frac{4\pi}{N}\sum_A
   \left.\left[
   \left(e^{\delta_A\partial_\xi}-1\right)
   \left(e^{\delta_A\partial_\eta}-1\right)
   I(\xi,\eta)
   +\left(e^{\delta_A\partial_\xi}-1\right)
   J(\xi)
   \right]\right|_{\xi=\eta=0}.
\label{eq:(3.23)}
\end{equation}
These representations show that the vacuum energy to the two-loop order is UV
finite, if the parameter~$\delta_A$ defined in~Eq.~\eqref{eq:(3.4)} is UV
finite. This implies that the parameter~$\delta\epsilon$ must receive a
non-trivial renormalization, as
\begin{equation}
   \text{$\delta_A=\frac{\lambda\delta\epsilon}{\pi RN}m_A$ is UV finite}
   \Rightarrow
   \delta\epsilon=\left(\frac{e^{\gamma_E}\mu^2}{4\pi}\right)^{-\varepsilon}
   \left(1+\frac{\lambda_R}{4\pi}\frac{1}{\varepsilon}\right)
   \delta\epsilon_R,
\label{eq:(3.24)}
\end{equation}
so that $\lambda\delta\epsilon=\lambda_R\delta\epsilon_R$ is UV finite; here we
have used~Eq.~\eqref{eq:(2.10)}.

In terms of the renormalized parameters, the expansion
of~Eq.~\eqref{eq:(3.22)} with respect to~$\delta\epsilon$ yields
\begin{align}
   \left.E^{(1)}\delta\epsilon\right|_{\text{1-loop}}
   &=N\Lambda\frac{1}{\Lambda R}
   \frac{\lambda_R\delta\epsilon_R}{\pi N}\frac{R}{N}
   \sum_Am_A\sum_{n\neq0}e^{-i(m_A+A_{y0})2\pi Rn}
   K_0(2\pi\Lambda R|n|),
\notag\\
   \left.E^{(2)}\delta\epsilon^2\right|_{\text{1-loop}}
   &=N\Lambda\frac{1}{(\Lambda R)^3}
   \left(\frac{\lambda_R\delta\epsilon_R}{\pi N}\right)^2
   \frac{R^2}{N}\sum_Am_A^2\left(-\frac{1}{4}\right)
\notag\\
   &\qquad{}
   \times\left[1+\sum_{n\neq0}e^{-i(m_A+A_{y0})2\pi Rn}
   2\pi\Lambda R|n|K_1(2\pi\Lambda R|n|)\right].
\label{eq:(3.25)}
\end{align}

For Eq.~\eqref{eq:(3.23)}, we need to carry out momentum integrations
in~Eqs.~\eqref{eq:(3.16)} and~\eqref{eq:(3.17)}. This is the subject of the
next subsection.

\subsection{$p$-integration in~$E^{(1)}\delta\epsilon|_{\text{2-loop}}$ and
$E^{(2)}\delta\epsilon^2|_{\text{2-loop}}$}
\label{sec:3.5}
Let us next consider $E^{(1)}\delta\epsilon|_{\text{2-loop}}$, which is given by
the $O(\delta_A)$ term of~Eq.~\eqref{eq:(3.23)}. By using the formulas
in~Appendix~\ref{sec:B}, $p$-integration in~Eq.~\eqref{eq:(3.17)} yields
\begin{align}
   \left.E^{(1)}\delta\epsilon\right|_{\text{2-loop}}
   &=2\pi R\frac{1}{N}\sum_A\delta_A
   \int\frac{d\ell_x}{2\pi}\frac{1}{2\pi R}\sum_{\ell_y}
   \left[
   \frac{\mathcal{L}(\ell)}{\mathcal{D}(\ell)}\Lambda^2
   +\frac{\mathcal{K}(\ell)^2}{\mathcal{D}(\ell)\mathcal{L}(\ell)}
   \frac{\Bar{\ell}_y^2}{\ell^2}
   \right]
\notag\\
   &\qquad{}
   \times\int_0^1dx\,
   \frac{1}{2}\sum_{n\neq0}e^{-i(m_A+A_{y0})2\pi Rn}e^{ix\ell_y2\pi Rn}
\notag\\
   &\qquad\qquad{}
   \times\Biggl\{
   (2\pi Rn)^2\left[K_0(z)-K_2(z)\right]\frac{2}{\ell^2}
   +(2\pi Rn)^2K_0(z)(-8)\frac{\ell_y^2}{(\ell^2)^2}
\notag\\
   &\qquad\qquad\qquad{}
   +\frac{2\pi R|n|}{\sqrt{x(1-x)\ell^2+\Lambda^2}}K_1(z)
   \left[\frac{4}{\ell^2}
   +i2\pi Rn\frac{\ell_y}{\ell^2}(-4)(1-2x)\right]   
   \Biggr\},
\label{eq:(3.26)}
\end{align}
where
\begin{equation}
   z\equiv\sqrt{x(1-x)\ell^2+\Lambda^2}2\pi R|n|.
\label{eq:(3.27)}
\end{equation}
Actually, the form of the integrand in the above expression depends on the
choice of the Feynman parameter~$x$. It can be changed by the change of
variables $x\to1-x$ and~$\ell_y\to-\ell_y$, which keeps the integration region
and the factor~$e^{ix\ell_y2\pi Rn}$ intact.\footnote{Recall that
$\ell_y\in\mathbb{Z}/R$.} It is convenient to fix the form of the
integrand~$\mathcal{I}(x,\ell_y)$ by
\begin{equation}
   \int_0^1dx\,\sum_{\ell_y}\mathcal{I}(x,\ell_y)
   \to\int_0^1dx\,\sum_{\ell_y}\frac{1}{2}
   \left[\mathcal{I}(x,\ell_y)+\mathcal{I}(1-x,-\ell_y)\right],
\label{eq:(3.28)}
\end{equation}
so that the form of the integrand is invariant under the above change of
variables. The particular expression in~Eq.~\eqref{eq:(3.26)} has been
obtained in this way.

Next, in Eq.~\eqref{eq:(3.26)} we use the identity
\begin{equation}
   K_{\nu-1}(z)-K_{\nu+1}(z)=-\frac{2\nu}{z}K_\nu(z)
\label{eq:(3.29)}
\end{equation}
with~$\nu=1$. Then, by further using
\begin{equation}
   K_0'(z)=-K_1(z)
\label{eq:(3.30)}
\end{equation}
and
\begin{equation}
   \frac{\partial z}{\partial x}
   =\frac{2\pi R|n|}{\sqrt{x(1-x)\ell^2+\Lambda^2}}(1-2x)\frac{\ell^2}{2},
\label{eq:(3.31)}
\end{equation}
which follows from~Eq.~\eqref{eq:(3.27)}, we have
\begin{align}
   \left.E^{(1)}\delta\epsilon\right|_{\text{2-loop}}
   &=2\pi R\frac{1}{N}\sum_A\delta_A
   \int\frac{d\ell_x}{2\pi}\frac{1}{2\pi R}\sum_{\ell_y}
   \left[
   \frac{\mathcal{L}(\ell)}{\mathcal{D}(\ell)}\Lambda^2
   +\frac{\mathcal{K}(\ell)^2}{\mathcal{D}(\ell)\mathcal{L}(\ell)}
   \frac{\Bar{\ell}_y^2}{\ell^2}
   \right]
\notag\\
   &\qquad{}
   \times\int_0^1dx\,
   \frac{1}{2}\sum_{n\neq0}e^{-i(m_A+A_{y0})2\pi Rn}e^{ix\ell_y2\pi Rn}
\notag\\
   &\qquad\qquad{}
   \times\left[
   2\pi Rn\ell_yK_0(z)
   -i\frac{\partial}{\partial x}K_0(z)
   \right]2\pi Rn(-8)\frac{\ell_y}{(\ell^2)^2}.
\label{eq:(3.32)}
\end{align}

Finally, integration by parts with respect to~$x$ yields
\begin{equation}
   \left.E^{(1)}\delta\epsilon\right|_{\text{2-loop}}=0.
\label{eq:(3.33)}
\end{equation}

Next, let us consider $E^{(2)}\delta\epsilon^2|_{\text{2-loop}}$, which is given
by the $O(\delta_A^2)$ terms in~Eq.~\eqref{eq:(3.23)}. First, the
$p$-integration in the function~$J$ in~Eq.~\eqref{eq:(3.17)} gives
\begin{align}
   &\left.E^{(2)}\delta\epsilon^2\right|_{\text{2-loop}}^{(J)}
\notag\\
   &=-2\pi R\frac{1}{N}\sum_A\delta_A^2
   \int\frac{d\ell_x}{2\pi}\frac{1}{2\pi R}\sum_{\ell_y}\int_0^1dx\,
   \left[
   \frac{\mathcal{L}(\ell)}{\mathcal{D}(\ell)}\Lambda^2
   +\frac{\mathcal{K}(\ell)^2}{\mathcal{D}(\ell)\mathcal{L}(\ell)}
   \frac{\Bar{\ell}_y^2}{\ell^2}
   \right]
\notag\\
   &\qquad{}
   \times\Biggl(
   \frac{1}{\left[x(1-x)\ell^2+\Lambda^2\right]^3}
   \left[-x(1-x)(3-10x+10x^2)-(1-2x+2x^2)\frac{\Lambda^2}{\ell^2}\right]
\notag\\
   &\qquad\qquad{}
   +\frac{1}{4}\sum_{n\neq0}e^{-i(m_A+A_{y0})2\pi Rn}e^{ix\ell_y2\pi Rn}
\notag\\
   &\qquad\qquad\qquad{}
   \times\Biggl\{
   \left(\frac{2\pi R|n|}
   {\sqrt{x(1-x)\ell^2+\Lambda^2}}\right)^3K_3(z)
\notag\\
   &\qquad\qquad\qquad\qquad\qquad{}
   \times
   \left[-2x(1-x)(1-3x+3x^2)
   -(1-2x+2x^2)\frac{\Lambda^2}{\ell^2}\right]
\notag\\
   &\qquad\qquad\qquad\qquad{}
   +\frac{(2\pi Rn)^2}{x(1-x)\ell^2+\Lambda^2}K_2(z)
\notag\\
   &\qquad\qquad\qquad\qquad\qquad{}
   \times
   \left[2(1-2x+2x^2)\frac{1}{\ell^2}
   +i2\pi Rn\frac{\ell_y}{\ell^2}(-2)(1-2x)(1-3x+3x^2)\right]
\notag\\
   &\qquad\qquad\qquad\qquad{}
   +\frac{(2\pi R|n|)^3}
   {\sqrt{x(1-x)\ell^2+\Lambda^2}}K_1(z)
\notag\\
   &\qquad\qquad\qquad\qquad\qquad{}
   \times
   \left[(1-2x+2x^2)\frac{1}{\ell^2}
   -4(1-2x+2x^2)\frac{\ell_y^2}{(\ell^2)^2}\right]   
   \Biggr\}
   \Biggr).
\label{eq:(3.34)}
\end{align}

On the other hand, from the function~$I$ in~Eq.~\eqref{eq:(3.16)},
\begin{align}
   &\left.E^{(2)}\delta\epsilon^2\right|_{\text{2-loop}}^{(I)}
\notag\\
   &=-2\pi R\frac{1}{N}\sum_A\delta_A^2
   \int\frac{d\ell_x}{2\pi}\frac{1}{2\pi R}\sum_{\ell_y}
   \int_0^1dx\,
\notag\\
   &\qquad{}
   \times
   \Biggl[
   \frac{\mathcal{L}(\ell)}{\mathcal{D}(\ell)}\ell^2
   \Biggl(
   \frac{1}{\left[x(1-x)\ell^2+\Lambda^2\right]^3}
   (-2)x(1-x)
   \left[
   x(1-x)-(1-6x+6x^2)\frac{\Lambda^2}{\ell^2}
   -2\frac{\Lambda^4}{(\ell^2)^2}
   \right]
\notag\\
   &\qquad\qquad{}
   +\frac{1}{4}\sum_{n\neq0}e^{-i(m_A+A_{y0})2\pi Rn}e^{ix\ell_y2\pi Rn}
\notag\\
   &\qquad\qquad\qquad{}
   \times\Biggl\{
   \left(\frac{2\pi R|n|}
   {\sqrt{x(1-x)\ell^2+\Lambda^2}}\right)^3K_3(z)
   (-2)x^2(1-x)^2
   \left(1+4\frac{\Lambda^2}{\ell^2}\right)
\notag\\
   &\qquad\qquad\qquad\qquad{}
   +\frac{(2\pi Rn)^2}{x(1-x)\ell^2+\Lambda^2}K_2(z)
\notag\\
   &\qquad\qquad\qquad\qquad\qquad{}
   \times 2x(1-x)
   \left\{
   2\frac{1}{\ell^2}
   +4\frac{\Lambda^2}{(\ell^2)^2}
   -i2\pi Rn\frac{\ell_y}{\ell^2}(1-2x)
   \left(1+4\frac{\Lambda^2}{\ell^2}\right)
   \right\}   
\notag\\
   &\qquad\qquad\qquad\qquad{}
   +\frac{(2\pi R|n|)^3}
   {\sqrt{x(1-x)\ell^2+\Lambda^2}}K_1(z)(-2)x(1-x)
   \left[
   \frac{1}{\ell^2}
   +4\Lambda^2\frac{\ell_y^2}{(\ell^2)^3}\right]   
   \Biggr\}
   \Biggr)
\notag\\
   &\qquad\qquad{}
   +\frac{\mathcal{K}(\ell)^2}{\mathcal{D}(\ell)\mathcal{L}(\ell)}
   \frac{\Bar{\ell}_y^2}{\ell^2}
   \Biggl(
   \frac{1}{\left[x(1-x)\ell^2+\Lambda^2\right]^3}4x(1-x)
   \left(1-3x+3x^2+\frac{\Lambda^2}{\ell^2}\right)
\notag\\
   &\qquad\qquad\qquad{}
   +\frac{1}{4}\sum_{n\neq0}e^{-i(m_A+A_{y0})2\pi Rn}e^{ix\ell_y2\pi Rn}
\notag\\
   &\qquad\qquad\qquad\qquad{}
   \times\Biggl\{
   \left(\frac{2\pi R|n|}
   {\sqrt{x(1-x)\ell^2+\Lambda^2}}\right)^3K_3(z)
   2x(1-x)(1-2x)^2
\notag\\
   &\qquad\qquad\qquad\qquad\qquad{}
   +\frac{(2\pi Rn)^2}{x(1-x)\ell^2+\Lambda^2}K_2(z)8x(1-x)
   \left[
   \frac{1}{\ell^2}
   -i2\pi Rn\frac{\ell_y}{\ell^2}(1-2x)
   \right]
\notag\\
   &\qquad\qquad\qquad\qquad\qquad{}
   +\frac{(2\pi R|n|)^3}
   {\sqrt{x(1-x)\ell^2+\Lambda^2}}K_1(z)
   (-8)x(1-x)\frac{\ell_y^2}{(\ell^2)^2}
   \Biggr\}
   \Biggr)
\notag\\
   &\qquad\qquad{}
   +\frac{\mathcal{K}(\ell)}{\mathcal{D}(\ell)}\frac{\Bar{\ell}_y^2}{\ell^2}
   \frac{1}{4}\sum_{n\neq0}e^{-i(m_A+A_{y0})2\pi Rn}e^{ix\ell_y2\pi Rn}
\notag\\
   &\qquad\qquad\qquad\qquad{}
   \times\
   \frac{(2\pi Rn)^2}{x(1-x)\ell^2+\Lambda^2}K_2(z)
   i2\pi Rn(-4)x(1-x)
   \Biggr].
\label{eq:(3.35)}
\end{align}
To obtain the expressions in~Eqs.~\eqref{eq:(3.34)} and~\eqref{eq:(3.35)}, we
applied the procedure in~Eq.~\eqref{eq:(3.28)}.

To further simplify the above expressions, we first note that all the terms
linear in~$\ell_y$ are proportional to~$1-2x$, and thus
to~$\partial z/\partial x$ as in~\eqref{eq:(3.31)}. Using this fact and the
identity
\begin{equation}
   K_2(z)=-z\left[\frac{1}{z}K_1(z)\right]',
\label{eq:(3.36)}
\end{equation}
we can carry out the integration by parts with respect to~$x$ in those terms
linear in~$\ell_y$. We then use the identity in~Eq.~\eqref{eq:(3.29)}
with~$\nu=2$ to express $K_3(z)$ in terms of~$K_1(z)$ and~$K_2(z)$. The
resulting expression contains the term~$K_1(z)x(1-x)(1-2x)^2$, for which we use
Eq.~\eqref{eq:(3.31)}. We repeat the integration by parts as long as the
factor~$1-2x$ remains. In an intermediate step, we use
\begin{equation}
   K_0(z)=-\frac{1}{z}\left[zK_1(z)\right]'.
\label{eq:(3.37)}
\end{equation}
Finally, we can carry out the $x$-integration in terms that do not contain the
Bessel function.\footnote{We note that
\begin{equation}
   \tanh^{-1}\left(\sqrt{\frac{\ell^2}{\ell^2+4\Lambda^2}}\right)
   =\frac{1}{4}\sqrt{\ell^2(\ell^2+4\Lambda^2)}\mathcal{L}_\infty(\ell).
\label{eq:(3.38)}
\end{equation}
}
In this way, we have the following rather simple expression:
\begin{align}
   &\left.E^{(2)}\delta\epsilon^2\right|_{\text{2-loop}}
\notag\\
   &=-2\pi R\frac{1}{N}\sum_A\delta_A^2
   \int\frac{d\ell_x}{2\pi}\frac{1}{2\pi R}\sum_{\ell_y}
\notag\\
   &\qquad{}
   \times\Biggr[
   \frac{\mathcal{L}(\ell)}{\mathcal{D}(\ell)}
   \frac{4-2(\ell^2+2\Lambda^2)\mathcal{L}_\infty(\ell)}
   {\ell^2(\ell^2+4\Lambda^2)}
\notag\\
   &\qquad\qquad{}
   +\int_0^1dx\,\sum_{n\neq0}e^{-i(m_A+A_{y0})2\pi Rn}e^{ix\ell_y2\pi Rn}
\notag\\
   &\qquad\qquad\qquad{}
   \times
   \Biggl(\frac{\mathcal{L}(\ell)}{\mathcal{D}(\ell)}
   \Biggl\{
   -\frac{(2\pi R|n|)^3}{\sqrt{x(1-x)\ell^2+\Lambda^2}}K_1(z)
   x(1-x)
\notag\\
   &\qquad\qquad\qquad\qquad\qquad\qquad{}
   -\frac{(2\pi Rn)^2}{x(1-x)\ell^2+\Lambda^2}K_2(z)
   x(1-x)
\notag\\
   &\qquad\qquad\qquad\qquad\qquad\qquad{}
   +\frac{\ell_y^2}{\ell^2}
   \Biggl[
   \frac{(2\pi R|n|)^3}{\sqrt{x(1-x)\ell^2+\Lambda^2}}K_1(z)
   x(1-x)
\notag\\
   &\qquad\qquad\qquad\qquad\qquad\qquad\qquad\qquad{}
   +(2\pi Rn)^2K_0(z)\frac{2}{\ell^2}
   \Biggr]
   \Biggr\}
\notag\\
   &\qquad\qquad\qquad\qquad{}
   -\frac{\mathcal{K}(\ell)}{\mathcal{D}(\ell)}
   \frac{\Bar{\ell}_y^2}{\ell^2}
   \frac{(2\pi Rn)^2}{x(1-x)\ell^2+\Lambda^2}K_2(z)
   i2\pi Rnx(1-x)
   \Biggr)
   \Biggr].
\label{eq:(3.39)}
\end{align}
This completes the $p$-integration in~$E^{(2)}\delta\epsilon^2|_{\text{2-loop}}$.

Let us examine whether the expression in~Eq.~\eqref{eq:(3.39)} is IR finite or
not. From the expressions in~Eq.~\eqref{eq:(2.18)} and
\begin{equation}
   \mathcal{L}_\infty(\ell)
   =\frac{1}{\Lambda^2}-\frac{1}{6}\frac{\ell^2}{\Lambda^2}+O((\ell^2)^2),   
\label{eq:(3.40)}
\end{equation}
we see that the above $\ell_x$-integral
for~$E^{(2)}\delta\epsilon^2|_{\text{2-loop}}$ is IR finite, as should be the case
for any physical quantity.

In what follows, we carry out the summation over the index~$A$
in~Eqs.~\eqref{eq:(3.25)} and~\eqref{eq:(3.39)} and integrate the resulting
expressions over the ``vacuum moduli''~$A_{y0}$ as in~Eq.~\eqref{eq:(2.11)}.
Then, we organize them according to the powers of~$1/N$. Before doing these,
however, it is helpful to further simplify Eq.~\eqref{eq:(3.39)} by noting that
$\Hat{\mathcal{L}}(p)$ and~$\mathcal{K}(p)$ in~Eqs.~\eqref{eq:(2.18)} are
exponentially suppressed for~$N\to\infty$ as~$\lesssim e^{-\Lambda RN}$ because
of the asymptotic behavior of the Bessel function,
$K_\nu(z)\stackrel{z\to\infty}{\sim}\sqrt{\pi/(2z)}e^{-z}$. Therefore, these
functions can be neglected in the power series expansion in~$1/N$ and we can
set $\mathcal{L}(\ell)\to\mathcal{L}_\infty(\ell)$,
$\mathcal{K}(\ell)\to0$,
and~$\mathcal{D}(\ell)\to(p^2+4\Lambda^2)\mathcal{L}_\infty(\ell)^2$
in~Eq.~\eqref{eq:(3.39)} to yield
\begin{align}
   &\left.E^{(2)}\delta\epsilon^2\right|_{\text{2-loop}}
\notag\\
   &=-2\pi R\frac{1}{N}\sum_A\delta_A^2
   \int\frac{d\ell_x}{2\pi}\frac{1}{2\pi R}\sum_{\ell_y}
\notag\\
   &\qquad{}
   \times\Biggr[
   \frac{4-2(\ell^2+2\Lambda^2)\mathcal{L}_\infty(\ell)}
   {\ell^2(\ell^2+4\Lambda^2)^2\mathcal{L}_\infty(\ell)}
\notag\\
   &\qquad\qquad{}
   +\int_0^1dx\,\sum_{n\neq0}e^{-i(m_A+A_{y0})2\pi Rn}e^{ix\ell_y2\pi Rn}
   \frac{1}{(\ell^2+4\Lambda^2)\mathcal{L}_\infty(\ell)}
\notag\\
   &\qquad\qquad\qquad{}
   \times
   \Biggl\{
   -\frac{(2\pi R|n|)^3}{\sqrt{x(1-x)\ell^2+\Lambda^2}}K_1(z)
   x(1-x)
   -\frac{(2\pi Rn)^2}{x(1-x)\ell^2+\Lambda^2}K_2(z)
   x(1-x)
\notag\\
   &\qquad\qquad\qquad\qquad{}
   +\frac{\ell_y^2}{\ell^2}
   \left[
   \frac{(2\pi R|n|)^3}{\sqrt{x(1-x)\ell^2+\Lambda^2}}K_1(z)
   x(1-x)
   +(2\pi Rn)^2K_0(z)\frac{2}{\ell^2}
   \right]
   \Biggr\}
   \Biggr],
\label{eq:(3.41)}
\end{align}
up to exponentially small terms.

\subsection{Summation over~$A$ and integration over~$A_{y0}$}
We thus consider the sum over the index~$A$ and the integration over the
vacuum moduli~$A_{y0}$ in~Eq.~\eqref{eq:(2.11)}. The summation over~$A$ can be
carried out as
\begin{equation}
   \sum_Ae^{-im_A2\pi Rn}=\sum_{j=0}^{N-1}\left(e^{-2\pi ni/N}\right)^j
   =N\begin{cases}
   1,&\text{for $n=0\bmod N$},\\
   0,&\text{for $n\neq0\bmod N$},\\
   \end{cases}
\label{eq:(3.42)}
\end{equation}
\begin{equation}
   \sum_Am_Ae^{-im_A2\pi Rn}
   =\frac{N}{2R}\begin{cases}
   1-\dfrac{1}{N},&\text{for $n=0\bmod N$},\\
   \dfrac{2}{N}\dfrac{1}{e^{-2\pi ni/N}-1},&\text{for $n\neq0\bmod N$},\\
   \end{cases}
\label{eq:(3.43)}
\end{equation}
and
\begin{equation}
   \sum_Am_A^2e^{-im_A2\pi Rn}
   =\frac{N}{3R^2}\begin{cases}
   1-\dfrac{3}{2N}+\dfrac{1}{2N^2},&\text{for $n=0\bmod N$},\\
   \dfrac{3}{N}\dfrac{1}{e^{-2\pi ni/N}-1}
   \left(1-\dfrac{2}{N}\dfrac{1}{1-e^{2\pi ni/N}}\right),
   &\text{for $n\neq0\bmod N$}.\\
   \end{cases}
\label{eq:(3.44)}
\end{equation}

On the other hand, the integration over~$A_{y0}$ with the measure
in~Eq.~\eqref{eq:(2.11)} results in
\begin{equation}
   \int_0^1d(A_{y0}RN)\,e^{-iA_{y0}2\pi Rn}
   =\begin{cases}
   1,&\text{for $n=0$},\\
   0,&\text{for $n\neq0$, $n=0\bmod N$},\\
   \dfrac{iN}{2\pi n}
   \left(e^{-2\pi ni/N}-1\right),&
   \text{for $n\neq0\bmod N$}.\\
   \end{cases}
\label{eq:(3.45)}
\end{equation}

The combination of the above two operations therefore yields
\begin{equation}
   \int_0^1d(A_{y0}RN)\,\sum_Am_A\,e^{-i(m_A+A_{y0})2\pi Rn}
   =\frac{N}{2R}\begin{cases}
   1-\dfrac{1}{N},&\text{for $n=0$},\\
   0,&\text{for $n\neq0$, $n=0\bmod N$},\\
   \dfrac{i}{\pi n},&
   \text{for $n\neq0\bmod N$},\\
   \end{cases}
\label{eq:(3.46)}
\end{equation}
and
\begin{align}
   &\int_0^1d(A_{y0}RN)\,\sum_Am_A^2\,e^{-i(m_A+A_{y0})2\pi Rn}
\notag\\
   &=\frac{N}{3R^2}\begin{cases}
   1-\dfrac{3}{2N}+\dfrac{1}{2N^2},&\text{for $n=0$},\\
   0,&\text{for $n\neq0$, $n=0\bmod N$},\\
   \dfrac{3i}{2\pi n}
   \left(1-\dfrac{1}{N}\right)
   +\dfrac{3}{2N}
   \dfrac{1}{\pi n}\dfrac{1}{\tan(\pi n/N)},&
   \text{for $n\neq0\bmod N$}.\\
   \end{cases}
\label{eq:(3.47)}
\end{align}

Using Eqs.~\eqref{eq:(3.46)} and~\eqref{eq:(3.47)} for~Eq.~\eqref{eq:(3.25)},
under the integration over~$A_{y0}$,
\begin{equation}
   \left.E^{(1)}\delta\epsilon\right|_{\text{1-loop}}
   =N\Lambda\frac{1}{\Lambda R}
   \frac{\lambda_R\delta\epsilon_R}{\pi N}
   \frac{1}{2}\sum_{n\neq0\bmod N}\frac{i}{\pi n}K_0(2\pi\Lambda R|n|)=0,
\label{eq:(3.48)}
\end{equation}
and
\begin{align}
   &\left.E^{(2)}\delta\epsilon^2\right|_{\text{1-loop}}
\notag\\
   &=N\Lambda\frac{1}{(\Lambda R)^3}
   \left(\frac{\lambda_R\delta\epsilon_R}{\pi N}\right)^2
   \left(-\frac{1}{12}\right)
   \left[
   1-\frac{3}{2N}+\frac{1}{2N^2}
   +\frac{6}{N}\sum_{n>0,n\neq0\bmod N}
   \frac{\Lambda RK_1(2\pi\Lambda Rn)}{\tan(\pi n/N)}
   \right].
\label{eq:(3.49)}
\end{align}

For the two-loop corrections, from~Eq.~\eqref{eq:(3.33)},
\begin{equation}
   \left.E^{(1)}\delta\epsilon\right|_{\text{2-loop}}=0,
\label{eq:(3.50)}
\end{equation}
and for Eq.~\eqref{eq:(3.40)} we have
\begin{align}
   &\left.E^{(2)}\delta\epsilon^2\right|_{\text{2-loop}}
\notag\\
   &=-\frac{2\pi}{3}\left(\frac{\lambda_R\delta\epsilon_R}{\pi RN}\right)^2
   \int\frac{d\ell_x}{2\pi}\frac{1}{2\pi R}\sum_{\ell_y}
\notag\\
   &\qquad{}
   \times\Biggr(
   \frac{1}{R}
   \left(1-\frac{3}{2N}+\frac{1}{2N^2}\right)
   \frac{4-2(\ell^2+2\Lambda^2)\mathcal{L}_\infty(\ell)}
   {\ell^2(\ell^2+4\Lambda^2)^2\mathcal{L}_\infty(\ell)}
\notag\\
   &\qquad\qquad{}
   +\int_0^1dx\,\sum_{n>0,n\neq0\bmod N}
\notag\\
   &\qquad\qquad\qquad{}
   \times
   \left[
   \frac{6}{N}\frac{\cos(x\ell_y2\pi Rn)}{\tan(\pi n/N)}
   -6\left(1-\frac{1}{N}\right)\sin(x\ell_y2\pi Rn)
   \right]
   \frac{1}{(\ell^2+4\Lambda^2)\mathcal{L}_\infty(\ell)}
\notag\\
   &\qquad\qquad\qquad\qquad{}
   \times
   \Biggl\{
   -\frac{(2\pi Rn)^2}{\sqrt{x(1-x)\ell^2+\Lambda^2}}K_1(z)
   x(1-x)
   -\frac{2\pi Rn}{x(1-x)\ell^2+\Lambda^2}K_2(z)
   x(1-x)
\notag\\
   &\qquad\qquad\qquad\qquad\qquad{}
   +\frac{\ell_y^2}{\ell^2}
   \left[
   \frac{(2\pi Rn)^2}{\sqrt{x(1-x)\ell^2+\Lambda^2}}K_1(z)
   x(1-x)
   +2\pi RnK_0(z)\frac{2}{\ell^2}
   \right]
   \Biggr\}
   \Biggr),
\label{eq:(3.51)}
\end{align}
up to exponentially small terms.

\subsection{Final results}
Finally, we arrange the above results in powers of~$1/N$. From
Eqs.~\eqref{eq:(3.48)} and~\eqref{eq:(3.50)}, we have
\begin{equation}
   E^{(1)}\delta\epsilon=0\cdot N^0+0\cdot N^{-1}+O(N^{-2}).
\label{eq:(3.52)}
\end{equation}
Thus, $E^{(1)}\delta\epsilon$ vanishes to the order we worked out.

For $E^{(2)}\delta\epsilon^2$, setting
\begin{equation}
   E^{(2)}\delta\epsilon^2
   =\left.E^{(2)}\delta\epsilon^2\right|_{O(N^{-1})}
   +\left.E^{(2)}\delta\epsilon^2\right|_{O(N^{-2})}
   +O(N^{-3}),
\label{eq:(3.53)}
\end{equation}
from~Eq.~\eqref{eq:(3.49)},
\begin{equation}
   \left.RE^{(2)}\delta\epsilon^2\right|_{O(N^{-1})}
   =N^{-1}(\lambda_R\delta\epsilon_R)^2(\Lambda R)^{-2}F(\Lambda R),
\label{eq:(3.54)}
\end{equation}
where
\begin{equation}
   F(\xi)\equiv
   -\frac{1}{12\pi^2}
   \left[1+c(\xi)\right],\qquad
   c(\xi)\equiv
   \lim_{N\to\infty}
   \frac{6}{N}\sum_{n>0,n\neq0\bmod N}
   \frac{\xi K_1(2\pi\xi n)}{\tan(\pi n/N)}.
\label{eq:(3.55)}
\end{equation}

From Eqs.~\eqref{eq:(3.49)} and~\eqref{eq:(3.51)}, on the other hand,
\begin{equation}
   \left.RE^{(2)}\delta\epsilon^2\right|_{O(N^{-2})}
   =N^{-2}(\lambda_R\delta\epsilon_R)^2(\Lambda R)^{-3}G(\Lambda R),
\label{eq:(3.56)}
\end{equation}
where
\begin{align}
   &G(\xi)
\notag\\
   &\equiv
   -\frac{1}{12\pi^2}
   \left\{
   -\frac{3}{2}\xi+
   \lim_{N\to\infty}\left[
   6\sum_{n>0,n\neq0\bmod N}
   \frac{\xi^2K_1(2\pi\xi n)}{\tan(\pi n/N)}-N\xi c(\xi)\right]
   \right\}
\notag\\
   &\qquad{}
   -\frac{1}{6\pi^3}
   \xi^3\int_{-\infty}^\infty d\Tilde{\ell}_x\,\sum_{\Tilde{\ell}_y\in\mathbb{Z}}
   \Biggr(
   \frac{4-2(\Tilde{\ell}^2+2\xi^2)\Tilde{\mathcal{L}}_\infty(\Tilde{\ell},\xi)}
   {\Tilde{\ell}^2(\Tilde{\ell}^2+4\xi^2)^2
   \Tilde{\mathcal{L}}_\infty(\Tilde{\ell},\xi)}
\notag\\
   &\qquad\qquad\qquad{}
   +\lim_{N\to\infty}\int_0^1dx\,\sum_{n>0,n\neq0\bmod N}
\notag\\
   &\qquad\qquad\qquad{}
   \times
   \left[
   \frac{6}{N}\frac{\cos(x\Tilde{\ell}_y2\pi n)}{\tan(\pi n/N)}
   -6\sin(x\Tilde{\ell}_y2\pi n)
   \right]
   \frac{1}{(\Tilde{\ell}^2+4\xi^2)\Tilde{\mathcal{L}}_\infty(\Tilde{\ell},\xi)}
\notag\\
   &\qquad\qquad\qquad\qquad{}
   \times
   \Biggl\{
   -\frac{(2\pi n)^2}{\sqrt{x(1-x)\Tilde{\ell}^2+\xi^2}}K_1(z)
   x(1-x)
\notag\\
   &\qquad\qquad\qquad\qquad\qquad{}
   -\frac{2\pi n}{x(1-x)\Tilde{\ell}^2+\xi^2}K_2(z)
   x(1-x)
\notag\\
   &\qquad\qquad\qquad\qquad\qquad{}
   +\frac{\Tilde{\ell}_y^2}{\Tilde{\ell}^2}
   \left[
   \frac{(2\pi n)^2}{\sqrt{x(1-x)\Tilde{\ell}^2+\xi^2}}K_1(z)
   x(1-x)
   +2\pi nK_0(z)\frac{2}{\Tilde{\ell}^2}
   \right]
   \Biggr\}
   \Biggr).
\label{eq:(3.57)}
\end{align}
In this expression, we have defined
\begin{equation}
   \Tilde{\mathcal{L}}_\infty(\Tilde{\ell},\xi)
   \equiv
   \frac{2}{\sqrt{\Tilde{\ell}^2(\Tilde{\ell}^2+4\xi^2)}}
   \ln\left(\frac{\sqrt{\Tilde{\ell}^2+4\xi^2}+\sqrt{\Tilde{\ell}^2}}
   {\sqrt{\Tilde{\ell}^2+4\xi^2}-\sqrt{\Tilde{\ell}^2}}\right)
\label{eq:(3.58)}
\end{equation}
and
\begin{equation}
   z\equiv\sqrt{x(1-x)\Tilde{\ell}^2+\xi^2}\,2\pi|n|.
\label{eq:(3.59)}
\end{equation}

\begin{figure}[htbp]
\centering
\includegraphics[width=0.6\columnwidth]{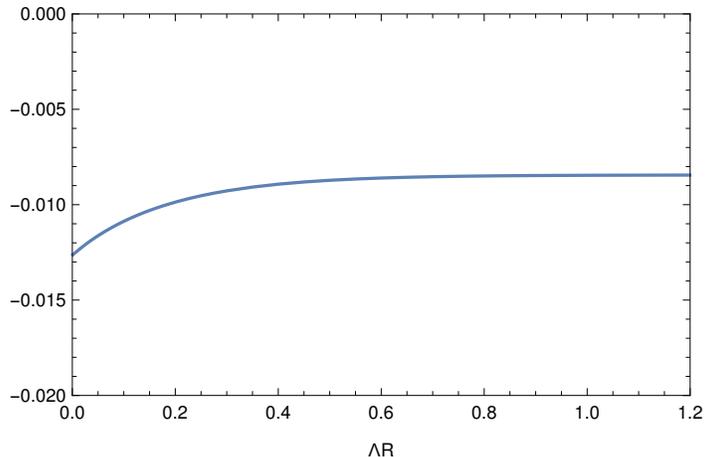}
\caption{The function~$F(\Lambda R)$ from~Eq.~\eqref{eq:(3.54)}.}
\label{fig:2}
\end{figure}

\begin{figure}[htbp]
\centering
\includegraphics[width=0.6\columnwidth]{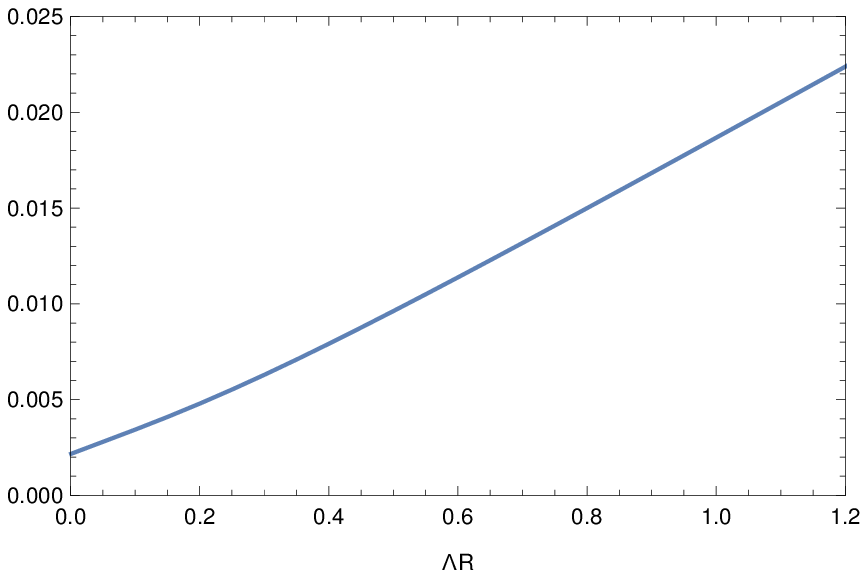}
\caption{The function~$G(\Lambda R)$ from~Eq.~\eqref{eq:(3.56)}.}
\label{fig:3}
\end{figure}

We plot the function~$F(\Lambda R)$ in~Eq.~\eqref{eq:(3.54)}
in~Fig.~\ref{fig:2} and the function~$G(\Lambda R)$ in~Eq.~\eqref{eq:(3.56)}
in~Fig.~\ref{fig:3}. These plots clearly show that, to the order of the $1/N$
expansion we worked out, the vacuum energy is a well-defined finite quantity
under the parameter renormalization in~Eqs.~\eqref{eq:(2.10)}
and~\eqref{eq:(3.24)}.
Equations~\eqref{eq:(3.52)}--\eqref{eq:(3.57)} and Figs.~\ref{fig:2}
and~\ref{fig:3} are the main results of this paper. Since Figs.~\ref{fig:2}
and~\ref{fig:3} show that the functions $F(\Lambda R)$ and~$G(\Lambda R)$
remain finite as~$\Lambda R\to0$, Eqs.~\eqref{eq:(3.54)} and~\eqref{eq:(3.56)}
[and Eq.~\eqref{eq:(3.52)}] show that the vacuum energy normalized by the
radius of the~$S^1$, $RE(\delta\epsilon)$, behaves as inverse powers
of~$\Lambda R$ for~$\Lambda R$ small, the $O(N^{-1})$ term behaves
as~$(\Lambda R)^{-2}$, and the $O(N^{-2})$ term behaves as~$(\Lambda R)^{-3}$.
Since $\Lambda$ is given by~Eq.~\eqref{eq:(2.9)}, this result implies that to
the order of the $1/N$ expansion we worked out, the vacuum energy is a purely
non-perturbative quantity and it has no well-defined weak coupling expansion
in~$\lambda_R$.

\section{Conclusion and discussion}
\label{sec:4}
By employing the $1/N$ expansion, we have computed the vacuum
energy~$E(\delta\epsilon)$ of the 2D SUSY $\mathbb{C}P^{N-1}$ model
on~$\mathbb{R}\times S^1$ with $\mathbb{Z}_N$ twisted boundary conditions to
the second order in the SUSY-breaking parameter~$\delta\epsilon$
in~Eq.~\eqref{eq:(3.2)}. We found that the vacuum energy is purely
non-perturbative and, although it is a perfectly well-defined physical
quantity in the $1/N$ expansion, it has no sensible weak coupling expansion
in~$\lambda_R$.

Our original intention was to compare our result in the $1/N$ expansion with
the result by the bion calculus in~Ref.~\cite{Fujimori:2018kqp}, because it
appears that the calculation in~Ref.~\cite{Fujimori:2018kqp} holds even under
the limit in~Eq.~\eqref{eq:(1.1)}.

According to~Ref.~\cite{Fujimori:2018kqp}, the contribution of a single bion
to the vacuum energy in~Eq.~\eqref{eq:(3.1)} is given by ($E^{(0)}$ is set to
be zero)
\begin{equation}
   RE^{(1)}\delta\epsilon
   =-R\sum_{b=1}^{N-1}2m_b\mathcal{A}_b(\Lambda R)^{2Rm_bN}\delta\epsilon
\label{eq:(4.1)}
\end{equation}
and
\begin{equation}
   RE^{(2)}\delta\epsilon^2
   =-R\sum_{b=1}^{N-1}2m_b\mathcal{A}_b(\Lambda R)^{2Rm_bN}
   \left[
   -2\gamma_E-2\ln\left(\frac{4\pi Rm_bN}{\lambda_R}\right)\mp\pi i\right]
   \delta\epsilon^2,
\label{eq:(4.2)}
\end{equation}
where the last $\mp\pi i$~term is the imaginary ambiguity caused by the
integration over quasi-collective coordinates of the bion. In these
expressions, the index~$b$ corresponds to the ``species'' of the bion and the
coefficient~$\mathcal{A}_b$ is given by using the twist angle~$m_A$
in~Eq.~\eqref{eq:(2.4)} as
\begin{align}
   \mathcal{A}_b
   &=\left[\frac{{\mit\Gamma}(1-m_bR)}{{\mit\Gamma}(1+m_bR)}\right]^2
   \prod_{a=1,a\neq b}^{N-1}\frac{m_a}{m_a-m_b}
   \frac{{\mit\Gamma}(1+(m_a-m_b)R)}{{\mit\Gamma}(1-(m_a-m_b)R)}
   \frac{{\mit\Gamma}(1-m_aR)}{{\mit\Gamma}(1+m_aR)}
\notag\\
   &=(-1)^{b+1}\frac{N^{2b}}{(b!)^2}.
\label{eq:(4.3)}
\end{align}
Using this, the coefficient of the imaginary ambiguity in~Eq.~\eqref{eq:(4.2)}
is given by
\begin{align}
   -R\sum_{b=1}^{N-1}2m_b\mathcal{A}_b(\Lambda R)^{2Rm_bN}
   =\frac{2}{N}\sum_{b=1}^{N-1}(-1)^b\frac{b}{(b!)^2}(\Lambda R N)^{2b}.
\label{eq:(4.4)}
\end{align}
When $N$ is fixed, in the weak coupling limit $\Lambda R\ll1$ for which the
semi-classical approximation should be valid, the $b=1$ term $-2\Lambda^2R^2N$
dominates the sum in~Eq.~\eqref{eq:(4.4)}. $\Lambda^2=\mu^2e^{-4\pi/\lambda_R}$ is
the exponential of the action of the constituent of the minimum bion (the
minimal fractional instanton--anti-instanton pair) and, at the same time, is
consistent with the order of the $u=1$ IR renormalon ambiguity. On the other
hand, in the large-$N$ limit in~Eq.~\eqref{eq:(1.1)}, whether
Eq.~\eqref{eq:(4.4)} possesses a sensible $1/N$ expansion or not is not clear,
because each term behaves as $O(N)$, $O(N^3)$, $O(N^5)$, \dots; we could not
estimate the sum as a whole in the large-$N$ limit.

Thus, we cannot compare our result in the $1/N$ expansion with the result
in~Ref.~\cite{Fujimori:2018kqp} by the bion calculus. We have no clear idea yet
why this comparison is impossible. One phenomenological observation
from~Eq.~\eqref{eq:(4.4)} is that it is a series in the
combination~$\Lambda RN$ and thus the result in~Ref.~\cite{Fujimori:2018kqp}
seems meaningful for~$\Lambda R N\ll1$ instead of our large-$N$ limit
in~Eq.~\eqref{eq:(1.1)}, with which $\Lambda R N\gg1$.\footnote{We would like
to thank Aleksey Cherman, Yuya Tanizaki, and Mithat \"Unsal for providing us
with suggestive arguments on this point.} More thought seems to be necessary to
clearly understand the relation between bions, the IR renormalon, and the $1/N$
expansion.

\section*{Acknowledgments}
We are grateful to Akira Nakayama and Hiromasa Takaura for collaboration at
various stages of this work. We would also like to thank Toshiaki Fujimori,
Tatsuhiro Misumi, Norisuke Sakai, and Kazuya Yonekura for helpful discussions.
This work was supported by JSPS Grants-in-Aid for Scientific Research numbers
JP18J20935 (O.M.) and~JP16H03982 (H.S.).

\section*{Note added}
In this paper we considered the large-$N$ limit specified
by~Eq.~\eqref{eq:(1.1)}, with which $N\Lambda R\to\infty$. On the other hand,
Ref.~\cite{Unsal:2008ch} discussed that the semi-classical picture such as that
in~Refs.~\cite{Argyres:2012vv,Argyres:2012ka,Dunne:2012ae,Dunne:2012zk} holds
only for~$N\Lambda R\ll1$. This is natural because the characteristic mass
scale with the twisted boundary condition can be $N\Lambda R$ instead
of~$\Lambda R$ and in the weak coupling limit $\Lambda\to0$. In this paper, we
also observed that the perturbative analyses cannot be available reasonably
for~$N\Lambda R\gg1$; our approximation is basically the expansion
in~$1/(N\Lambda R)$ and it is impossible to read how the vacuum energy behaves
as~$N\Lambda R\to0$ from our large-$N$ result. In a recent
paper~\cite{Morikawa:2020agf}, perturbation theory with the twisted boundary
condition is carefully studied for~$N\Lambda R\to0$ and a picture consistent
with the bion calculus has been obtained.

\appendix

\section{The perturbative part of the vacuum energy contains IR divergences}
\label{sec:A}
In the limit~$R\to\infty$, the expression of the vacuum energy is considerably
simplified because $n\neq0$ terms  in~Eqs.~\eqref{eq:(3.55)}
and~\eqref{eq:(3.57)} are exponentially suppressed in this limit. We have
\begin{equation}
   RE^{(2)}\delta\epsilon^2
   \stackrel{R\to\infty}{\to}
   -\frac{1}{12\pi^2}(\lambda_R\delta\epsilon_R)^2
   \left\{
   N^{-1}(\Lambda R)^{-2}
   +N^{-2}\left[-\frac{3}{2}(\Lambda R)^{-2}+G_{\infty}\right]
   +O(N^{-3})
   \right\},
\label{eq:(A1)}
\end{equation}
where
\begin{equation}
   G_\infty\equiv
   \frac{8\pi}{R^2}\int\frac{d\ell_x}{2\pi}\frac{1}{2\pi R}
   \sum_{\ell_y}
   \frac{4-2(\ell^2+2\Lambda^2)\mathcal{L}_\infty(\ell)}
   {\ell^2(\ell^2+4\Lambda^2)^2\mathcal{L}_\infty(\ell)}.
\end{equation}
Equation~\eqref{eq:(A1)} is a non-perturbative expression obtained to the
next-to-leading order of the $1/N$ expansion. From Eq.~\eqref{eq:(3.40)}, we
see that the $\ell$-integration in~$G_\infty$ is IR convergent.

To extract the perturbative part from~Eq.~\eqref{eq:(A1)}, we expand~$G_\infty$
with respect to~$\Lambda$ and neglect all terms with positive powers
of~$\Lambda=\mu e^{-2\pi/\lambda_R}$. Noting the behavior
$\mathcal{L}_\infty\sim(2/\ell^2)\ln(\ell^2/\Lambda^2)$
from~Eq.~\eqref{eq:(2.18)}, we obtain the perturbative part as
\begin{equation}
   G_\infty\sim
   \frac{8\pi}{R^2}\int\frac{d\ell_x}{2\pi}\frac{1}{2\pi R}
   \sum_{\ell_y}
   \frac{2}{(\ell^2)^2}
   \left[\frac{1}{\ln(\ell^2/\Lambda^2)}-1\right].
\label{eq:(A3)}
\end{equation}
The perturbative expansion with respect to~$\lambda_R(\mu)$ is then given by
\begin{equation}
   G_\infty\sim
   \frac{8\pi}{R^2}\int\frac{d\ell_x}{2\pi}\frac{1}{2\pi R}
   \sum_{\ell_y}
   \frac{2}{(\ell^2)^2}
   \left[-1+\sum_{k=0}^\infty[-\ln(\ell^2/\mu^2)]^k
   \left(\frac{\lambda_R}{4\pi}\right)^{k+1}\right],
\label{eq:(A4)}
\end{equation}
where we have used
\begin{equation}
   \ln(\ell^2/\Lambda^2)=\ln(\ell^2/\mu^2)
   +\frac{4\pi}{\lambda_R(\mu)}.
\end{equation}
Equations~\eqref{eq:(A3)} and~\eqref{eq:(A4)} show that the perturbative part
of~$G_\infty$ suffers from IR divergences in the $\ell$-integration, although
the full $G_\infty$ itself is IR finite.

\section{Integration formulas}
\label{sec:B}
In Sect.~\ref{sec:3.5} we have used the following integration formulas (in
practice, we are interested in the cases $(\alpha,\beta)=(1,2)$, $(1,3)$,
and~$(2,2)$):
\begin{align}
   &\int\frac{d^2p}{(2\pi)^2}\,e^{ip_y2\pi Rn}\,
   \frac{1}{\left[(p-\ell)^2+\Lambda^2\right]^\alpha}
   \frac{1}{(p^2+\Lambda^2)^\beta}
   \begin{cases}
   1\\
   p_\mu\\
   p_\mu p_\nu\\
   \end{cases}
\notag\\
   &\stackrel{n=0}{=}
   \frac{1}{{\mit\Gamma}(\alpha){\mit\Gamma}(\beta)}
   \int_0^1dx\,x^{\alpha-1}(1-x)^{\beta-1}
\notag\\
   &\qquad{}\times
   \frac{1}{4\pi}
   \begin{cases}
   {\mit\Gamma}(\alpha+\beta-1)
   \left[x(1-x)\ell^2+\Lambda^2\right]^{1-\alpha-\beta}.\\
   {\mit\Gamma}(\alpha+\beta-1)
   \left[x(1-x)\ell^2+\Lambda^2\right]^{1-\alpha-\beta}x\ell_\mu,\\
   {\mit\Gamma}(\alpha+\beta-1)
   \left[x(1-x)\ell^2+\Lambda^2\right]^{1-\alpha-\beta}x^2\ell_\mu\ell_\nu\\
   \qquad{}+\frac{1}{2}{\mit\Gamma}(\alpha+\beta-2)
   \left[x(1-x)\ell^2+\Lambda^2\right]^{2-\alpha-\beta}\delta_{\mu\nu},\\
   \end{cases}
\notag\\
   &\stackrel{n\neq0}{=}
   \frac{1}{{\mit\Gamma}(\alpha){\mit\Gamma}(\beta)}
   \int_0^1dx\,x^{\alpha-1}(1-x)^{\beta-1}
\notag\\
   &\qquad{}\times
   \frac{1}{4\pi}2^{2-\alpha-\beta}e^{ix\ell_y2\pi Rn}
   \begin{cases}
   \left(\frac{2\pi R|n|}{\sqrt{x(1-x)\ell^2+\Lambda^2}}
   \right)^{\alpha+\beta-1}
   K_{\alpha+\beta-1}(z),\\
   \left(\frac{2\pi R|n|}{\sqrt{x(1-x)\ell^2+\Lambda^2}}
   \right)^{\alpha+\beta-1}
   K_{\alpha+\beta-1}(z)x\ell_\mu\\
   \qquad{}+
   \left(\frac{2\pi R|n|}{\sqrt{x(1-x)\ell^2+\Lambda^2}}
   \right)^{\alpha+\beta-2}
   K_{\alpha+\beta-2}(z)i2\pi Rn\delta_{\mu y},\\
   \left(\frac{2\pi R|n|}{\sqrt{x(1-x)\ell^2+\Lambda^2}}
   \right)^{\alpha+\beta-1}
   K_{\alpha+\beta-1}(z)x^2\ell_\mu\ell_\nu\\
   \qquad{}+\left(\frac{2\pi R|n|}{\sqrt{x(1-x)\ell^2+\Lambda^2}}
   \right)^{\alpha+\beta-2}
   K_{\alpha+\beta-2}(z)\\
   \qquad\qquad\qquad{}
   \times
   (\delta_{\mu\nu}+ix\ell_\mu2\pi Rn\delta_{\nu y}+i2\pi Rn\delta_{\mu y}x\ell_\nu)\\
   \qquad\qquad{}-\left(\frac{2\pi R|n|}{\sqrt{x(1-x)\ell^2+\Lambda^2}}
   \right)^{\alpha+\beta-3}
   K_{\alpha+\beta-3}(z)\\
   \qquad\qquad\qquad\qquad{}
   \times(2\pi Rn)^2\delta_{\mu y}\delta_{\nu y},\\
   \end{cases}
\end{align}
where
\begin{equation}
   z\equiv\sqrt{x(1-x)\ell^2+\Lambda^2}2\pi R|n|.
\end{equation}

\end{document}